\documentclass[twocolumn,natbib,tighten]{aastex62}

\pdfoutput=1

\usepackage{xspace}
\usepackage{float}
\usepackage{xcolor}
\usepackage{multirow}

\defcitealias{Mathewson1992}{M92}
\defcitealias{Mathewson1996}{M96}
\defcitealias{Courteau1997}{C97}
\defcitealias{Courteau1993}{C93}
\defcitealias{Courteau2000}{SF}
\defcitealias{Lelli2016}{L16}
\defcitealias{Ouellette2017}{SV}
\defcitealias{Lelli2017}{L17}
\defcitealias{McGaugh2016}{M16}

\newcommand{\Sec}[1]{{\protect\hyperref[sec:#1]{Sec.~\ref*{sec:#1}}}}
\newcommand{\Secs}[2]{{\protect\hyperref[sec:#1]{Secs.~\ref*{sec:#1}}~and~\ref{sec:#2}}}
\newcommand{\Fig}[1]{{\protect\hyperref[fig:#1]{Fig.~\ref*{fig:#1}}}}
\newcommand{\Equ}[1]{{\protect\hyperref[equ:#1]{Eq.~\ref*{equ:#1}}}}
\newcommand{\Equs}[2]{{\protect\hyperref[equ:#1]{Eqs.~\ref*{equ:#1}}~and~\ref{equ:#2}}}
\newcommand{\Tab}[1]{{\protect\hyperref[tab:#1]{Table~\ref*{tab:#1}}}}
\newcommand{\App}[1]{{\protect\hyperref[app:#1]{Appendix~\ref*{app:#1}}}}
\newcommand{\be}{\begin{equation}}                                                             
\newcommand{\ee}{\end{equation}}  

\newcommand{\HI}{H{\footnotesize I}\xspace}

\newcommand{\Ha}{H\ensuremath{\alpha}\xspace}
\newcommand{\gb}{\ensuremath{\rm{g}_{\rm{bar}}}\xspace}
\newcommand{\gs}{\ensuremath{\rm{g}_{\rm{*}}}\xspace}
\newcommand{\gob}{\ensuremath{\rm{g}_{\rm{obs}}}\xspace}
\newcommand{\gd}{\ensuremath{\rm{g}_{\dagger}}\xspace}

\newcommand{\Lst}{\citetalias{Lelli2016}\xspace}

\newcommand{\Courteau}{\citetalias{Courteau1997}\xspace}
\newcommand{\Cnt}{\citetalias{Courteau1993}\xspace}
\newcommand{\MATtwo}{\citetalias{Mathewson1992}\xspace}
\newcommand{\MATsix}{\citetalias{Mathewson1996}\xspace}
\newcommand{\Lseventeen}{\citetalias{Lelli2017}\xspace}
\newcommand{\Msixteen}{\citetalias{McGaugh2016}\xspace}

\newcommand{\sigmaobs}{0.17\xspace}
\newcommand{\sigmasim}{0.12\xspace}
\newcommand{\sigmaint}{0.11\xspace}

\newcommand{\wunits}[2]{\ensuremath{#1\,\text{#2}}}

\received{2019 April 24}
\revised{2019 June 27}
\accepted{2019 July 9}
\submitjournal{ApJ}

\shorttitle{Scatter of the Radial Acceleration Relation}
\shortauthors{Stone and Courteau}

\begin{document}

\title{The Intrinsic Scatter of the Radial Acceleration Relation\footnote{Released on \today}}

\correspondingauthor{Connor Stone}
\email{connor.stone@queensu.ca}

\author[0000-0002-9086-6398]{Connor Stone}
\affiliation{Department of Physics, Engineering Physics and Astronomy,
  Queen{'}s University,
  Kingston, ON K7L 3N6, Canada}

\author[0000-0002-8597-6277]{St{\'e}phane Courteau}
\affiliation{Department of Physics, Engineering Physics and Astronomy,
  Queen{'}s University,
  Kingston, ON K7L 3N6, Canada}

\begin{abstract}
  We present a detailed Monte Carlo model of observational errors in observed galaxy scaling relations to recover the intrinsic (cosmic) scatter driven by galaxy formation and evolution processes.
  We apply our method to the stellar radial acceleration relation (RAR) which compares the local observed radial acceleration to the local Newtonian radial acceleration computed from the stellar mass distribution.
  The stellar and baryonic RAR are known to exhibit similar scatter.  
  Lelli+2017 (L17) studied the baryonic RAR using a sample of 153 spiral galaxies and inferred a negligible intrinsic scatter.
  If true, a small scatter might challenge the $\Lambda$CDM galaxy formation paradigm, possibly favoring a modified Newtonian dynamics interpretation.
  The intrinsic scatter of the baryonic RAR is predicted by modern $\Lambda$CDM simulations to be $\sim$0.06-0.08 dex, contrasting with the null value reported by L17.
  We have assembled a catalog of structural properties with over 2500 spiral galaxies from six deep imaging and spectroscopic surveys (called PROBES for the ``Photometry and Rotation curve OBservations from Extragalactic Surveys'') to quantify the intrinsic scatter of the stellar RAR and other scaling relations.
  The stellar RAR for our full sample  has a median observed scatter of \sigmaobs dex.
  We use our Monte Carlo method, which accounts for all major sources of measurement uncertainty, to infer a contribution of \sigmasim dex from the observational errors.
  The intrinsic scatter of the stellar RAR is thus estimated to be $\sigmaint \pm 0.02$ dex, in agreement with, though slightly greater than, current $\Lambda$CDM predictions.
\end{abstract}


\section{Introduction} \label{sec:intro}

The correlation between two observables resulting from a physical causation is often referred to as a scaling relation.
In the context of galaxies, these relations may be driven by formation and evolution processes, often revealing a subtle interplay between the baryonic and dark matter components.
The slope, normalization, and scatter of these scaling relations are sensitive to galaxy formation parameters and mechanisms such as star formation efficiencies, merger histories, and the coupling between baryons and dark matter.
For instance, the velocity-luminosity relation and fundamental planes of late- and early-type galaxies can be reproduced through a fine balance between gravitational forces and radiative processes~\citep{Courteau2007,Dutton2007,Dutton2011,Trujillo-Gomez2011}.

The scatter of an observed relation is the sum of observational and intrinsic (cosmic) components. 
Observational uncertainties arise from a variety of measurement limitations, including low signals, foreground/background interlopers, imprecise distance indicators, fuzzy parameter definitions, ill-defined data quality control, selection biases, and more.
Conversely, the intrinsic scatter results from physical processes that have shaped galaxies since formation~\citep{Dutton2007,Somerville2015}.
Intrinsic scatter can be readily compared with predictions from numerical simulations of galaxies in order to discriminate viable formation models. 
The accurate assessment of the intrinsic scatter of scaling relations is thus of utmost value for galaxy formation studies.

In this paper, we present a technique to extract the intrinsic scatter of a scaling relation by constructing a detailed model of the observational uncertainties.
Doing so requires following a few operational steps.
Firstly, a scaling relation is constructed with observational data.
Two ideally uncorrelated variables are paired, and a global relation is inferred via an unbiased fitting method (see \Sec{fittingoobservedrar}).
Second, all observed data points are projected onto the fitted scaling relation.
This provides new values for the core measurements under an assumption of zero intrinsic scatter.
Third, the measurements are then resampled from their observational uncertainty distributions, centered on the new zero-scatter values.
Fourth, the scaling relation including only observational errors is constructed using the resampled data points.
It is now possible to directly compare the observed scaling relation with the one containing only observational errors, allowing one to infer the intrinsic variations.

We demonstrate the efficacy of our method by applying it to the radial acceleration relation (RAR)~\citep[][hereafter \Msixteen]{McGaugh2016}.  
The latter consists of recasting the velocity–mass relation, often referred to as the baryonic Tully–Fisher relation (BTFR), of galaxies into a form that involves centripetal accelerations.
While the BTFR uses global quantities, the RAR is constructed from spatially resolved parameters.
Specifically, the observed centripetal acceleration from spectroscopic measurements is compared with that inferred from the baryonic matter distribution inferred via photometry.
As our database does not include gas masses, we focus on the stellar RAR which exhibits the same scatter as the baryonic RAR.
The scatter of the stellar and baryonic RARs has been shown to be comparable for both Spitzer Photometry and Accurate Rotation Curves (SPARC) observations (\Lseventeen) and NIHAO simulations (A. Dutton 2019, private communication).
Discrepancies between the observed and expected radial accelerations of galaxies have typically been ascribed to a dark matter particle component~\citep{Bertone2005,Courteau2014,Bertone2018b}. 

\citet[][hereafter \Lseventeen]{Lelli2017} examined the scatter of the residuals in the baryonic RAR ($\sigma_{\rm{RAR}}$) for a heterogeneous compilation of 153 spiral galaxies, labeled ``SPARC,'' and found it to be $\sigma_{\rm{RAR}} = 0.13\,\rm{dex}$.
They also found that a Gaussian fit to the residuals returns $\sigma_{\rm{RAR}} = 0.11\,\rm{dex}$.
This indicates that the data are not normally distributed, otherwise the two scatter measurements would be equal.
A first-order analysis, assuming Gaussianity, of the observational uncertainties yields $\sigma_{\rm{obs}} \approx 0.12\,\rm{dex}$.
As this estimate is close to their total observed scatter value, \Lseventeen claimed that the baryonic RAR is consistent with zero intrinsic scatter.
Moreover, since $\Lambda$CDM theory predicts a nonzero intrinsic scatter about the relation \citet{Ludlow2017,Keller2017,Dutton2019}; \Lseventeen suggested that their inferred null scatter result favors Modified Newtonian Dynamics~\citep[MOND;][]{Milgrom1983}. 

Given the widespread success of $\Lambda$CDM, the claim by \Lseventeen remains controversial and calls for further scrutiny.
\citet{Navarro2017} showed that the baryonic RAR arises naturally from $\Lambda$CDM using a standard galaxy formation model.
The EAGLE, MUGS2, and NIHAO $\Lambda$CDM simulations also yield estimates for the intrinsic baryonic RAR scatter of $\sigma_{\rm{int},\Lambda\rm{CDM}} \approx 0.06-0.08\,\rm{dex}$ \citep{Ludlow2017,Keller2017,Dutton2019}.

It was pointed out by \Lseventeen and \citet{Wheeler2018} that a galaxy that obeys the BTFR ~\citep{Walker1999,McGaugh2000,Brook2016} typically obeys the baryonic RAR, except at the extremes of large and small radii not easily accessible with observations. 
When comparing the BTFR and RAR, the choice of axes for each relation is also relevant.
The BTFR is normally depicted as the stellar mass (larger relative error) on the Y-axis and the circular velocity (smaller relative error) on the X-axis~\citep{McGaugh2000,McGaugh2012,Ponomareva2018,Lelli2019}.
The corresponding representation for the RAR therefore calls for the inferred acceleration, \gb (larger relative error), on the Y-axis and the measured acceleration, \gob (small relative error), on the X-axis.  
The opposite approach was adopted by \Lseventeen. 
Their choice of axes for the RAR thus naturally favors a small forward scatter, $\Delta Y(X)$.
Indeed, the inverse scatter of the baryonic RAR, $\Delta X(Y)$, for the SPARC sample is nearly twice as large, on average, as the forward scatter (see \App{inversescatter}).
The ratio $\Delta Y(X)/\Delta X(Y)$ grows with radius for the SPARC sample, as the relation flattens to a slope of 0.5 for small \gb.
As we also show in \Sec{reintroducingobservationaluncertainties}, the largest contribution to the errors  is the uncertainty of the stellar mass-to-light ratio, $M_*/L$.
A forward scatter measurement thus calls for the variables $M_*$ and \gb on the Y-axis for the BTFR and RAR, respectively; this is discussed further in \App{inversescatter}.

On a related note, \citet{Rodrigues2018} found that the individual galaxies in the SPARC sample each favor slightly different acceleration scales for \gd indicating that its value is not universal, which MOND requires.
Clearly, this is a case where the intrinsic scatter of a scaling relation bears immediate consequences for the interpretation of its physical underpinnings.

In this paper, we examine the scatter of the stellar RAR in detail, using a statistically compelling collection of resolved photometric and kinematic profiles for spiral galaxies coupled with a comprehensive error analysis.
Whereas simulations seem to converge on $\sigma_{\rm{int},\Lambda\rm{CDM}} \approx 0.06-0.08\,\rm{dex}$, our approach consists of inferring $\sigma_{\rm{int},\rm{universe}}$ through careful elimination of all observational errors from the observed scatter.

In \Sec{therar}, we briefly review the stellar RAR and discuss the various sources of error that contribute to its overall scatter.
This is followed in \Sec{datasets} by a description of six large galaxy surveys, collectively referred to as the Photometry and Rotation curve OBservations from Extragalactic Surveys (PROBES) catalog, providing structural parameters for 2500 galaxies to be used for our extensive RAR analysis. 
Our description of each galaxy survey includes salient features that are relevant for the stellar RAR.
\Sec{uncertaintymodel} addresses the dominant sources of uncertainty in the stellar RAR and how these are modeled for our analysis, whereas \Sec{montecarloscattermodel} describes the technique used to determine the observational error contribution to the stellar RAR scatter.
In \Sec{results}, the stellar RAR scatter is computed for each survey using a full Monte Carlo uncertainty model.
Here the intrinsic scatter, $\sigma_{\rm{int}}$, is assessed by comparing the observed total scatter and the observational uncertainties in quadrature.
The value of $\sigma_{\rm{int}}$ is then compared with estimates from numerical galaxy formation models, as well as values obtained by \Lseventeen.
Our measurement of $\sigma_{\rm{int}} $ serves as an empirical validation of $\Lambda$CDM models. 
For simplicity, unless otherwise stated, ``RAR''  means ``stellar RAR'' throughout. The ``baryonic RAR'' will be explicitly stated when needed. 

\section{The RAR}
\label{sec:therar}

First reported in \Msixteen, the RAR is a tight relationship between the observed spatially resolved radial acceleration profile of a galaxy and that expected from baryonic matter alone.
The RAR is essentially a translation of the mass discrepancy acceleration relation \citep[MDAR;][]{McGaugh2004}. 
The RAR is also closely related to the BTFR, as the two can be shown to match if $V(r)$ is constant and $M_{\rm bar}(r)\approx M_{\rm bar, total}$, which is typically where the BTFR is constructed~\citep{Lelli2017,Wheeler2018}.
As stated in \Sec{intro}, the BTFR is typically represented with baryonic mass on the Y-axis, while \Lseventeen's RAR has the baryon-dependent quantity on the X-axis, hence contributing to the remarkably tight forward scatter, $Y(X)$,  observed by \Lseventeen as discussed in \Sec{intro}. 

The baryonic mass distribution (and resulting acceleration) of a galaxy is primarily composed of two parts, stellar and gaseous, such that $\gb = \gs + \rm{g}_{\rm{gas}}$, where $\gb$ is the observed baryonic radial acceleration, and $\gs$ and $\rm{g}_{\rm{gas}}$ are the contributions to $\gb$ from the stars and gas.
Save for the SPARC data set, the other PROBES surveys (\Sec{datasets}) all lack neutral gas measurements.  
For those, a stellar RAR (\gob versus \gs) can be readily constructed.
The scatters in the stellar and baryonic RAR are comparable, as seen in \Lseventeen observations and NIHAO simulations (A. Dutton 2019, private communication). The stellar RAR is thus also a strong constraint on galaxy formation and dark matter models, but with the benefit of being more easily accessible observationally than the baryonic RAR.

The baryonic RAR was parametrized by \Msixteen with a MOND-inspired single-parameter function.
We present the equation from \Msixteen in \Equ{rar}, now using \gs instead of \gb, as we will be examining the stellar RAR,

\begin{equation}\label{equ:rar}
  \gob = \frac{\gs}{1 - e^{-\sqrt{\gs/\gd}}}~,
\end{equation}

\noindent where \gs is the stellar radial acceleration, \gob is the observed radial acceleration, and \gd is the expected universal MONDian acceleration scale equal to approximately \wunits{10^{-10}}{m\,s$^{-2}$}.
This MOND-motivated fitting function has a slope of unity (1:1) for large \gs commonly found in the central regions of a galaxy.
In the limit that \gs is small, where standard galaxy formation models typically ascribe high dark matter fractions, that function becomes $\gob = \sqrt{\gd\gs}$ .
Thus, the representation of the RAR in \Equ{rar} is an alternative representation of MOND~\citep{Milgrom1983}.

The scatter of the RAR, or any other scaling relation, can be decomposed as the sum in quadrature of its intrinsic and observed components: $\sigma_{\rm{RAR}}^2 = \sigma_{\rm{int}}^2 + \sigma_{\rm{obs}}^2$, where $\sigma_{\rm{RAR}}$ is the scatter of the scaling relation residuals, $\sigma_{\rm{int}}$ is the intrinsic scatter of the relation, and $\sigma_{\rm{obs}}$ is the scatter due to observational uncertainties/errors.
The observational uncertainties of the RAR can be approximated to first order as: $\sigma_{\rm{obs}}^2 \approx \sigma_{\rm 1^{st} Order}^2 = \sigma_{\gob}^2 + \left(\frac{\partial\mathcal{F}}{\partial\gs}\sigma_{\gs}\right)^2$, where $\sigma_{\gob}$ is the observational uncertainty in \gob, and $\left(\frac{\partial\mathcal{F}}{\partial\gs}\sigma_{\gs}\right)^2$ is the observational uncertainty in \gs modulated by the functional form of the RAR relation (\Equ{rar}).
The values $\sigma_{\gob}$ and $\sigma_{\gs}$ can be further broken down based on the variables used in the calculation of \gs and \gob.
As shown below, numerous variables enter this calculation.
A first-order uncertainty propagation is also presented.

\subsection{Observed Radial Acceleration}
\label{sec:gobs}

The local observed radial acceleration \gob is computed from a galaxy rotation curve (RC) with the formula

\begin{equation}\label{equ:calcgobs}
    \gob = \frac{\left(V_{\rm{obs}}/\sin(i)\right)^2}{\theta D}~,
\end{equation}

\noindent where $V_{\rm{obs}}$ is the measured line-of-sight velocity, $\theta$ is the angular radius of the velocity measurement, $D$ is the distance to the galaxy, and the inclination $i$ is inferred from the axial ratios of the galaxy.
The latter, which represents the galaxy's photometric tilt relative to the line of sight, is computed according to

\begin{equation}\label{equ:calcinclination}
    \cos^2(i) = \frac{q^2 - q_{0}^2}{1 - q_{0}^2}~,
\end{equation}

\noindent where $q=b/a$ is the measured axis ratio of the semi-major, $a$, and semi-minor, $b$, axes of the isophote, and $q_{0} = h_{Z}/h_{R}$ is a parameter representing the intrinsic flattening of a galaxy (the ratio of disk scale height, $h_{Z}$, to scale length, $h_{R}$).
If $q < q_0$, an inclination of $90^{\circ}$ is used.
Typical values for $q_{0}$ are in the range 0.1 to 0.25~\citep{Kregel2002,Dutton2005,Hall2012}.
While the value of $q_{0}$ for a given galaxy cannot be measured directly, some correlations with other galaxy parameters do exist.
For instance, $h_{z}$ scales with $V_{max}$ in edge-on galaxies~\citep{Kregel2005}; however $V_{max}$ is an inclination- (and thus $q_{0}$-) dependent quantity and ill-suited for this analysis.
Here $q_{0}$ also correlates with morphological type (T), and we use three flattening formulations based on T-Type, $q_{0}^{[1-3]} = 0.20\pm0.03$, $q_{0}^{[4]} = 0.17\pm0.03$, and $q_{0}^{[5-10]} = 0.12\pm0.02$ where the superscript represents the T-Type~\citep{Haynes1984,deGrijs1998}.

The uncertainty for each introduced variable plays a significant role in determining $\sigma_{\gob}$, except $\sigma_{\theta}$ which is assumed to be negligibly small.
Distance uncertainties enter into the calculation of \gob only via $D$, as the velocity and inclination measurements are distance-independent.
Using \Equs{calcgobs}{calcinclination}, one may compute the first-order uncertainty on \gob,

\begin{equation}\label{equ:uncertaintygobs1storder}
  \left[\frac{\sigma_{\gob}}{\gob}\right]^2 = \left[\frac{2\sigma_{V_{\rm{obs}}}}{V_{\rm{obs}}}\right]^2 + \left[\frac{2q\sigma_{q}}{1-q^2}\right]^2 + \left[\frac{2q_0\sigma_{q_0}}{1-q_0^2}\right]^2 + \left[\frac{\sigma_{\rm{D}}}{\rm{D}}\right]^2~,
\end{equation}

\noindent where $\sigma_x$ is the observational uncertainty on variable $x$.
While this is formally the uncertainty on a single \gob measurement, it includes $D$ and $i$ which are shared variables between all \gob values for a given galaxy.
As the first order calculation does not account for shared variables, it cannot yield accurate predictions.
However, this simplified calculation is still useful for comparison with a full Monte Carlo model (\Sec{montecarloscattermodel}), and the analysis performed in \Lseventeen.
Depending on the relative values of $\sigma_{V_{\rm{obs}}}$ versus $\sigma_{\rm{D}}$ and $\sigma_{\rm{i}}$, $\sigma_{\gob}$ could be dominated by the local velocity measurement or shared galaxy variables ($D$ and $i$).
Once again, the RAR data are highly correlated, and a first-order error analysis will incorrectly predict $\sigma_{\rm{obs}}$.

\subsection{Stellar Radial Acceleration}
\label{sec:gbar}

The stellar radial acceleration, \gs, requires the specification of a three-dimensional spatial model for the stars, as the observed photometric values only yield a projected mass distribution.
In the simplest case, the stellar mass can be assumed to lie in spherical shells, and, for a given curve of growth, \gs can be represented as

\begin{equation}\label{equ:calcgbarspherical}
  \gs = \frac{G\Upsilon_{x} L_x}{(\theta D)^2} = \frac{G\Upsilon_{x} 10^{-(m_{x} - M_{\odot,x})/2.5}}{\theta^2}~,
\end{equation}

\noindent where $G$ is the gravitational constant, $\Upsilon_{x}$ is the stellar mass-to-light ratio in a photometric band $x$ in solar units, and $L_x$ and $m_{x}$ are the luminosity and apparent magnitude in band $x$ respectively, enclosed by an isophote at angular radius $\theta$.
Here, $M_{\odot,x}$ is the absolute magnitude of the Sun in band $x$.
The quantity \gs is fortunately independent of distance errors, as the distance dependence of the luminosity $L_x$ is canceled by the acceleration formula. 
The mass-to-light ratio, $\Upsilon_{x}$, is either constant (for the \wunits{3.6}{$\mu$m} band) or determined by a colour mass-to-light ratio formula and thus independent of distance.
The gravitational constant $G$ and photometric normalization $M_{\odot,x}$ are considered to have negligible error.
However, $\Upsilon_{x}$ can have considerable uncertainty~\citep{Conroy2013,Courteau2014,Roediger2015}, and the uncertainty on $m_{x}$ can be significant, especially for fainter galaxies.
Using \Equ{calcgbarspherical}, the first-order uncertainty on \gs is computed to be

\begin{equation}\label{equ:uncertaintygbar1storder}
    \left[\frac{\sigma_{\gs}}{\gs}\right]^2 = \left[\frac{\ln(10)}{2.5}\sigma_{m_{x}}\right]^2 + \left[\frac{\ln(10)}{2.5}\sigma_{m_{0}}\right]^2 + \left[\frac{\sigma_{\Upsilon}}{\Upsilon}\right]^2~.
\end{equation}

\noindent As in the \gob equation (\Equ{calcgobs}), some variables have shared uncertainty at all points in a single galaxy.
For instance, all isophotal magnitudes share the same photometric zero-point $m_{0}$ and its uncertainty.
Also, $m_{x}$ is correlated with all magnitude estimates interior to it, and assuming a constant $\Upsilon_{x}$ means that its uncertainty is shared across all points in the galaxy.
This means, like \gob, that the photometric data points \gs are not independent, and a first-order analysis will incorrectly predict $\sigma_{\gs}$.

\Equ{uncertaintygbar1storder} is generated under the assumption that a galaxy's mass distribution is spherical.
In reality, disk galaxies are flattened structures whose potential is calculated by solving Poisson's equation, $\nabla^2\phi = 4\pi G\rho$, where $\rho$ is the three-dimensional mass density.
The acceleration at each location on the disk is then computed as $\gs = -\nabla\phi$.
The value of \gs at a given radius depends strongly on the full mass distribution.
Therefore, individual \gs values for disk galaxies will be even more correlated than in the spherical case.
In our analysis, we use the three-dimensional density distribution from \citet{vanderKruit1981},

\begin{equation}
  \rho_{\rm{disk}}(R, Z) = \Sigma(R)\frac{\rm{sech}^2(Z/Z_{0})}{2Z_{0}}~,
\end{equation}

\noindent where $\Sigma(R)$ is the radial projected surface mass density, $Z$ is the height above the disk, and $Z_{0}$ is the sech scale height of the disk.
Here $h_{Z}$ can be determined from $q_0$ and $h_R$ by definition ($h_Z = q_0h_R$; see \Sec{gobs}) and we use $Z_{0} = 2h_{Z}$ to find the sech scale height~\citep{Kregel2002}.
With a density distribution specified, Poisson's equation is solved by \emph{galpy} \citep{Bovy2015} using a self-consistent field method~\citep{Hernquist1992}.
This treatment is only applied to observed galaxies and thus the observed RAR. For the Monte Carlo model described in \Sec{montecarloscattermodel}, the mock galaxies are treated as spherically symmetric, given the difficulty in inverting Poisson's equation.
In the next section, we discuss the galaxy samples used to construct and analyze the RAR. 

\section{Data sets}
\label{sec:datasets}

Our study relies on the compilation of six distinct surveys of 2634 disk galaxies, including the SPARC dataset of 163 galaxies assembled by \citet{Lelli2016} from various heterogeneous small datasets (see \Sec{datasetsSP} below).
The full sample, collectively referred to as PROBES will be presented elsewhere (C. Stone et~al. 2019, in preparation).
The six surveys presented below were collated in order to overcome small sample limitations, such as SPARC used on its own, and to mitigate selection biases.
For instance, with only 163 galaxies (albeit with extended \HI RCs),  SPARC suffers from small number statistics.  
Numerous SPARC galaxies were also handpicked, leaving open the potential for unintentional bias.  
Our description of the six samples below highlights various features, such as selection criteria, data quality cuts, and distance estimates, that may either benefit or hinder the determination of an unbiased RAR.
The maximal extent of available surface brightness (SB) profiles and RCs are reported below in terms of $R_{23.5}$, or the isophotal radius corresponding to \wunits{23.5}{mag\,arcsec$^{-2}$}, in the relevant photometric band.

Each selected survey includes spatially resolved light profiles, in at least one photometric band, as well as spatially resolved RCs usually extracted from \Ha long-slit spectra or \HI synthesis maps.
If only one photometric band is available for the spatially resolved light profiles, stellar mass-to-light ratios are recovered via global colors estimated by the authors or retrieved from the NASA Extragalactic Database (NED).
Spatially resolved \HI fluxes to compute gas masses at all galactocentric radii and, ultimately, baryonic masses are only available for the SPARC data set.
Therefore,  mass computations include only stellar masses inferred via suitably chosen $M_*/L$ transformations~\citep{Roediger2015,Zhang2017}.
Missing gas masses in the RAR analysis affect the relation normalization but not its scatter, which is central to this paper, as discussed briefly in \Sec{gbar}. 

\subsection{Our Survey Collection}
\label{sec:surveys}

\subsubsection{M92}

\citet[][hereafter \MATtwo]{Mathewson1992} acquired \Ha RCs and $I$-band photometry for 744 galaxies, mostly in the southern hemisphere.
Each galaxy has a measured global B-I color, except for 51 galaxies where color information was retrieved from NED.
Uncertainties for spatially resolved quantities are not reported for the SB profiles or RCs.
Instead, we use simple models, described in \Sec{uncertaintymodel}, to estimate reasonable uncertainties.
Repeat measurements by \MATtwo yielded typical SB errors less than \wunits{\sim0.05}{mag\,arcsec$^{-2}$} and velocity errors less than \wunits{10}{km\,s$^{-1}$}.
The median SB profile extends to 1.2$R_{23.5}$, and the median RC profile extends to 0.9$R_{23.5}$.
Galaxies were selected for this survey primarily from the ESO-Uppsala catalog \citep{lauberts1982, lauberts1998} with morphological types of Sb to Sd, diameters greater than $1.7'$, radial velocities typically below \wunits{7000}{km$\,$s$^{-1}$}, inclination above $40^{\circ}$, latitude $|b|$ greater than $11^{\circ}$, and a small number of galaxies from other surveys~(\MATtwo).

\subsubsection{M96}

\citet[][hereafter \MATsix]{Mathewson1996} extended the \MATtwo sample with 1216 additional galaxies with \Ha RCs and $I$-band photometry, primarily in the southern hemisphere.
Many of these galaxies do not have B-I global colors and so global colors were once again retrieved from NED for a total of 399 galaxies.
Similar to \MATtwo, uncertainties for spatially resolved quantities are not provided, and the models described in \Sec{uncertaintymodel} are used.
The sampling criteria were similar to those of \MATtwo, except with radial velocities in the range 4000 to \wunits{14,000}{km\,s$^{-1}$}, and apparent diameters between $1.0'$ and $1.6'$, again with a small number taken from other surveys, such as the Uppsala General Catalog.
The median SB profile extends to 1.2$R_{23.5}$, and the median RC profile extends to 0.8$R_{23.5}$.

\subsubsection{C97}

The \citet[][hereafter \Courteau]{Courteau1997} sample is a collection of 296 Sb- and Sc- type galaxies with \Ha RCs and $r$-band photometry.
These were collected largely for cosmic flow studies for which systematic and random uncertainties were of great interest \citep[][hereafter \Cnt]{Courteau1993}. 
Thus, many galaxies have repeat measurements, with some having as many as four remeasured RCs.
More than half of the galaxies have multiple SB profiles.
For this analysis, wherever multiple integrations exist, the deepest RC or SB profile is used.
Repeat measurements are still valuable for the purpose of assessing uncertainties.
The median SB profile uncertainty is \wunits{0.03}{mag\,arcsec$^{-2}$}, and the median RC profile uncertainty is \wunits{6}{km\,s$^{-1}$}.
The median SB profile extends to 1.3$R_{23.5}$, and the median RC profile extends to 1.0$R_{23.5}$.
The sample was selected from the Uppsala Catalog of Galaxies \citep{Nilson1973,lauberts1998}, and the catalog of cluster galaxies from \citet{Bothun1985}.
The galaxies were selected to have Hubble types Sb-Sc, Zwicky magnitude $m_{B} \leq 15.5$, blue galactic extinction less than \wunits{0.5}{mag} (based on \citealt{Burstein1984}), inclinations between $55^{\circ}$ and $75^{\circ}$, blue major axes less than $4'$, and to be noninteracting/merging and have no overlapping bright stars~\citep{Courteau1996}.

\subsubsection{\citet{Courteau2000}}

Shellflow, from \citet{Courteau2000}, is a sample of 171 galaxies with \Ha RCs and both $V$- and $I$-band photometry.
The survey was designed to study an all-sky shell in redshift space to measure a cosmological bulk flow of galaxies with high precision.
The Shellflow sample geometry meant that a large fraction of the galaxies could be observed from both the northern and southern hemisphere observatories, namely, KPNO and CTIO, thus mitigating calibration errors from using different instrumentation.  
The multiband photometry enables radially resolved mass-to-light ratios. 
The median SB profile uncertainty is \wunits{0.04}{mag\,arcsec$^{-2}$} and the median RC profile uncertainty is \wunits{6}{km\,s$^{-1}$}.
The full distributions can be found in \Secs{uncertaintymagvalues}{uncertaintyvelocities}.
The median SB profile extends to 1.4$R_{23.5}$, and the median RC profile extends to 0.9$R_{23.5}$.
The Shellflow sample was selected from the Optical Redshift Survey by \citet{Santiago1995}.
Galaxies were chosen to be noninteracting and of morphological types of Sb and Sc, with radial velocities between 4500 and \wunits{7000}{km\,s$^{-1}$}, inclinations between $45^{\circ}$ and $78^{\circ}$, $A_B$ extinctions less than \wunits{0.3}{mag} (as determined by \citealt{Burstein1982}), and having no bright overlapping foreground stars or tidal disturbances.

\subsubsection{L16}\label{sec:datasetsSP}

The Spitzer Photometry and Accurate Rotation Curves (SPARC) sample compiled by \citet[][hereafter \Lst]{Lelli2016} is an amalgamation of over 50 smaller samples totaling 163 galaxies\footnote{an additional 12 ``low quality'' galaxies are not used in this paper} with Spitzer \wunits{3.6}{$\mu$m} photometry and \HI RCs.
Approximately one-third of the SPARC galaxies have hybrid \HI and \Ha RCs, to combine the higher spatial resolution provided by \Ha, with the extensive radial extent of synthesis \HI radio maps, where available.
The distances to SPARC galaxies rely on a number of methods, including Hubble flow, tip of the red giant branch, Cepheids, Ursa Major cluster distance, and supernovae.
All distances and their uncertainties are included in the survey; we use them directly for our analysis.
The median SB profile uncertainty is \wunits{0.01}{mag\,arcsec$^{-2}$} and the median RC profile uncertainty is \wunits{4.0}{km\,s$^{-1}$}.
The full velocity uncertainty distribution can be found in \Sec{uncertaintyvelocities}; resolved magnitude uncertainties are not reported for SPARC and are therefore lacking in \Sec{uncertaintymagvalues}.
The median SB profile extends to 0.7$R_{23.5}$, and the median RC profile extends to 1.2$R_{23.5}$.
The sample was carefully chosen to have a wide range of morphology, luminosity, and SB from the limited selection of available galaxies with \HI measurements~(\Lst).
The photometry is homogeneously collected in the Spitzer \wunits{3.6}{$\mu$m} band, RCs are compiled from 56 separate studies~(\Lst) excluding THINGS~\citep{deBlok2008} and LITTLE-THINGS~\citep{Oh2015}.
Each RC in SPARC is assigned a quality flag, with $Q=1$ and $2$ considered acceptable and $Q=3$ (12 objects) having strong non-circular motions and/or asymmetric features, making them unsuitable for the analysis in \Lseventeen. 

\subsubsection{\citet{Ouellette2017}}\label{sec:datasetsSV}

The Spectroscopy and $H$-band Imaging of Virgo Cluster Galaxies (SHIVir) survey presented in \citet{Ouellette2017} is a dedicated survey of galaxies in the Virgo Cluster with dynamical and multiband information. 
The sample was selected to examine the impact of the cluster environment on galaxy properties using the nearest cluster in the sky.
For our RAR analysis, we focus on the subset of 44 SHIVir spiral galaxies with \Ha RCs from long-slit spectra and ugriz-H photometry. 
While SHIVir is the smallest survey in our RAR analysis, its spatially resolved multiband imaging makes it valuable for studying the importance of resolved versus global mass-to-light ratios, as mass-to-light ratios are a substantial source of uncertainty in the RAR.
Distances to the SHIVir galaxies are measured using SB fluctuations where available~\citep{Jerjen2004,Blakeslee2009}; otherwise a standard value of \wunits{16.5}{Mpc} is assumed~\citep{Mei2007}.
Uncertainties are reported where distance measurements are available; for the rest, a median value of \wunits{3}{Mpc} is used.
The median SB profile uncertainty is \wunits{0.03}{mag\,arcsec$^{-2}$} and the median RC profile uncertainty is \wunits{4}{km\,s$^{-1}$}, the full distributions can be seen in \Secs{uncertaintymagvalues}{uncertaintyvelocities}.
The median SB profile extends to 1.4$R_{23.5}$, and the median RC profile extends to 0.6$R_{23.5}$.
The SHIVir galaxies are a subset of 286 galaxies drawn from the Virgo Cluster Catalog (VCC), based on the intersection of the VCC catalog and SDSS 6th Data Release, which is volume complete in a spatial subset of the Virgo Cluster to an absolute magnitude of $M_B\leq -15.15$, along with several fainter galaxies, to ensure a broad morphology coverage~\citep{McDonald2011}.

\subsection{Data Quality Selections}
\label{sec:dataquality}

A number of data quality criteria were implemented in our compilation.
Only galaxies with inclinations greater than $30^{\circ}$ were considered; rotational velocities are corrected by $\sin(i)^{-1}$ and would therefore diverge at lower inclination.
This study focuses solely on late-type galaxies, and only galaxies with morphological types from Sa to Im are considered (T-Type of 1 - 10). 
Only a few other types were available in each survey and were not beneficial to this analysis.
Individual RC data points were discarded if $\sigma_V/V > 0.1$, as in \Lseventeen.
Individual SB data points were discarded if $\sigma_{\mu} > \wunits{0.1}{mag\,arcsec$^{-2}$}$ or $\mu > 25$.
Data points within $5''$ of the center of a galaxy were also discarded to avoid seeing effects.

\subsection{The RAR for Each Survey}
\label{sec:fittingoobservedrar}

We now present the RAR for each survey in \Fig{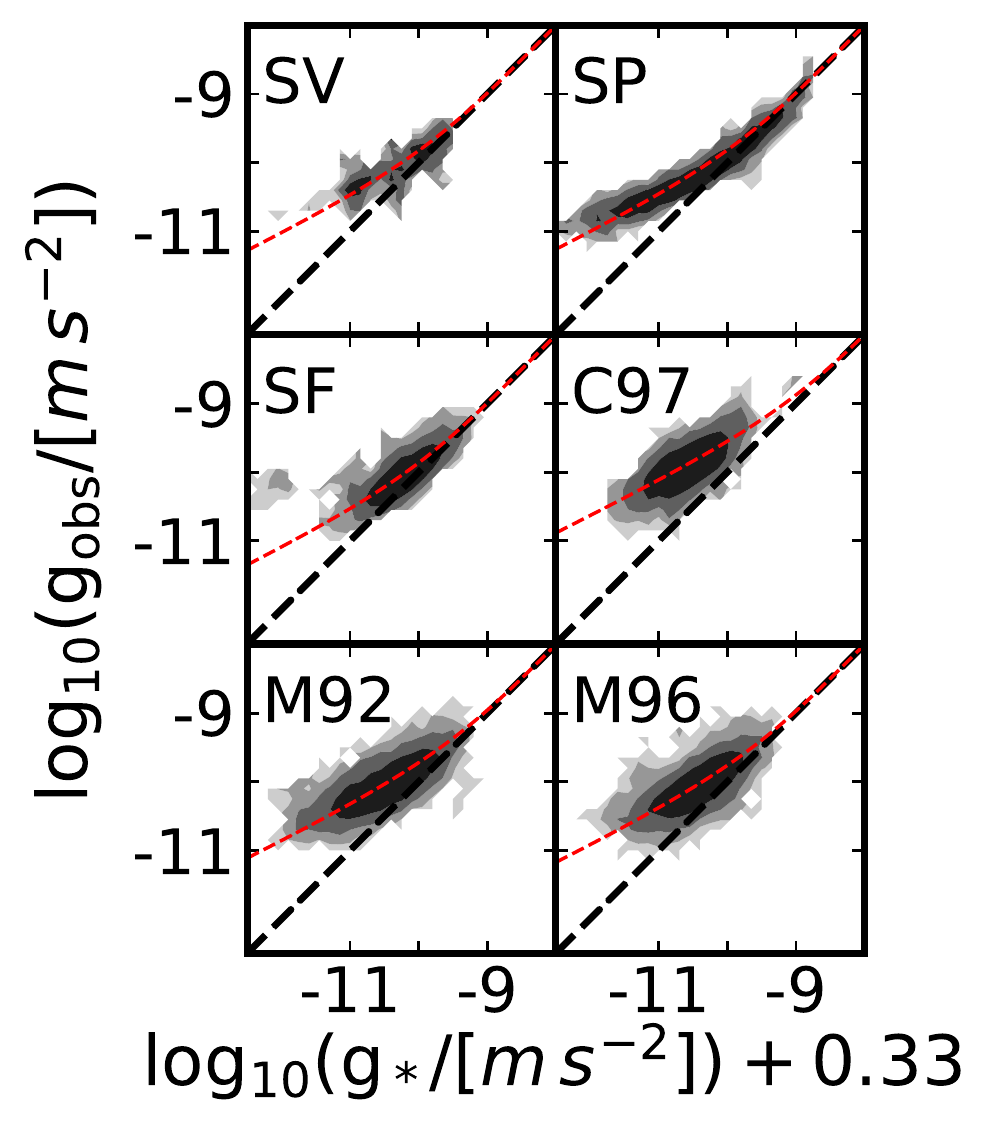}.
Each panel shows a fit to the data using \Equ{rar}; the fit uses an orthogonal distance regression ~\citep{scipy}.
The axes are calculated using the analysis described in \Secs{gobs}{gbar} plus an extra \wunits{0.33}{dex} added to the \gs axis to account for the missing gas mass to the measured stellar mass values. 
This is done largely for aesthetic and fitting reasons (properly positioned and reasonable fitted \gd values) and plays no role in our study of the RAR scatter.
The RAR scatter, $\sigma_{\rm RAR}$, is measured relative to a running median in a window chosen to include $\sim100$ datapoints.  
A parametric assessment of the scatter would be biased by the arbitrariness of the RAR functional form and suffer from an artificial increase to the scatter due to any misalignment with the data.
It is thus avoided here.

\begin{figure}[ht]
  \centering
  \includegraphics[width=\columnwidth]{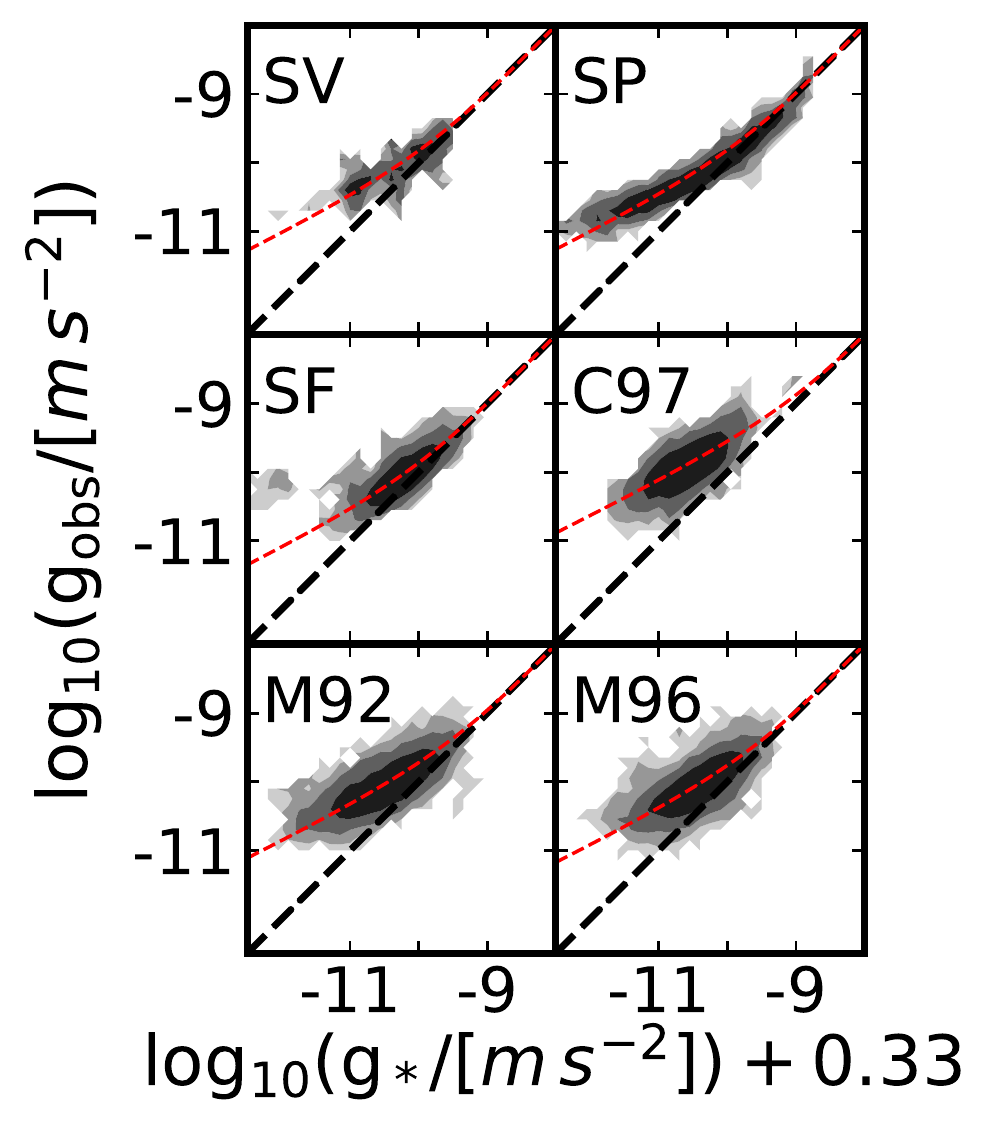}
  \caption{\footnotesize The RAR for all surveys presented as 2D contour plots with levels evenly spaced in log density.
    The black and red dashed lines represent the one-to-one line and the fit to the data from \Equ{rar}, respectively.
    Each panel shows a different survey indicated in the upper left corner. The surveys are abbreviated as SHIVir (SV), SPARC (SP), Shellflow (SF), \citealt{Courteau1997} (C97), \citealt{Mathewson1992} (M92), and \citealt{Mathewson1996} (M96).}
  \label{fig:Figures/rar_plot_obs.pdf}
\end{figure}

The values for \gd in \Tab{observedfits} differ by many standard deviations.
It is, however, important to note that that table only includes bootstrap random errors. 
The large systematic uncertainties that affect variables such as the mass-to-light ratios, which could exceed a factor of 2, are not included in this treatment.
Accounting for them would easily reconcile all of the values of \gd with each other~\citep{Courteau2014,Roediger2015}.
Therefore, strong statements regarding the universality of \gd based on our six surveys cannot be made.
As in \Fig{Figures/rar_plot_obs.pdf} and \Tab{observedfits}, each survey is analyzed separately throughout the paper.
Systematic differences between the surveys that could artificially inflate the observed scatter are thus minimized.

\begin{table}
  \footnotesize
  \begin{center}
    \caption{\\Parametrization of the RAR Fit to Each PROBES Data set}
    \begin{tabular}{ c  c  c  c  c }
      \tableline\tableline
      1 & 2 & 3 & 4 \\
      Survey & $\gd$ & $\sigma_{\rm RAR}$ & $N$ \\
      & ($10^{-10}~m\,s^{-1}$) & (dex) & \# \\ \tableline
      SV   & $0.813 \pm 0.044$ & $0.166 \pm 0.007$ & 1380 \\
      SP   & $0.839 \pm 0.020$ & $0.131 \pm 0.003$ & 2561 \\
      SF   & $0.558 \pm 0.008$ & $0.139 \pm 0.001$ & 16,648 \\
      C97  & $5.224 \pm 0.042$ & $0.182 \pm 0.001$ & 17,626 \\
      M92  & $1.790 \pm 0.016$ & $0.173 \pm 0.001$ & 18,355 \\
      M96  & $1.345 \pm 0.012$ & $0.170 \pm 0.001$ & 21,460 \\
      \tableline
    \end{tabular}\label{tab:observedfits}
  \end{center}
  \par {Note. Column (1) indicates the survey (as in \Fig{Figures/rar_plot_obs.pdf}). In column (2), \gd and its bootstrap uncertainty emerge from \Equ{rar}.  Column (3) gives the \wunits{16-84}{\%} interval scatter of the forward residuals, $\sigma_{\rm{RAR}}$, and its bootstrap uncertainty. Column (4) gives the number of data points in the RAR.}
\end{table}

\section{Uncertainty Model}
\label{sec:uncertaintymodel}

Many parameters enter the computation of the stellar and observed radial accelerations, and they must each be characterized when adopting an appropriate uncertainty model.
Once an error function is determined for each parameter, it can be sampled randomly to form the basis of the Monte Carlo model described in \Sec{montecarloscattermodel}.  
We examine each major parameter and its related uncertainty below. 

\subsection{Distances}
\label{sec:uncertaintydistances}

Depending on the method, distances may carry significant uncertainties and represent an important source of scatter in the RAR.
Most galaxies in the SHIVir and SPARC surveys are close enough to get distance measurements with variable stars, SB fluctuations, cluster distance, tip of the red giant branch, or supernovae.
Their different uncertainties are reported in each survey.
For the more distant galaxies in \MATtwo, \MATsix, \Courteau, and Shellflow, Hubble flow distances are used ~\citep{Reiss2016}.
A primary source of distance uncertainty when using this method is due to peculiar motions.
Here we assume the peculiar motion to be of order $\sigma_{pec} = \wunits{300}{km\,s$^{-1}$}$, where $\sigma_{pec}$ is the peculiar velocity dispersion.
Many galaxies in the surveys are far enough that this $\sigma_{pec}$ value is relatively small, so a lower relative uncertainty bound of $\sigma_D = 0.15D$ is applied.
Thus, for distance uncertainties, we have $\sigma_D = max(0.15D,~300)$ for Hubble flow distances.

\subsection{Inclinations}
\label{sec:uncertaintyinclinations}

Inclination uncertainties are modeled in multiple steps.
First, a per-galaxy axis ratio uncertainty is computed using the scatter in isophotal ellipticity of the outer regions of the SB profile (ellipticities beyond $R_{e}$).
These cover a large range of isophotal fitting uncertainties, as shown in \Fig{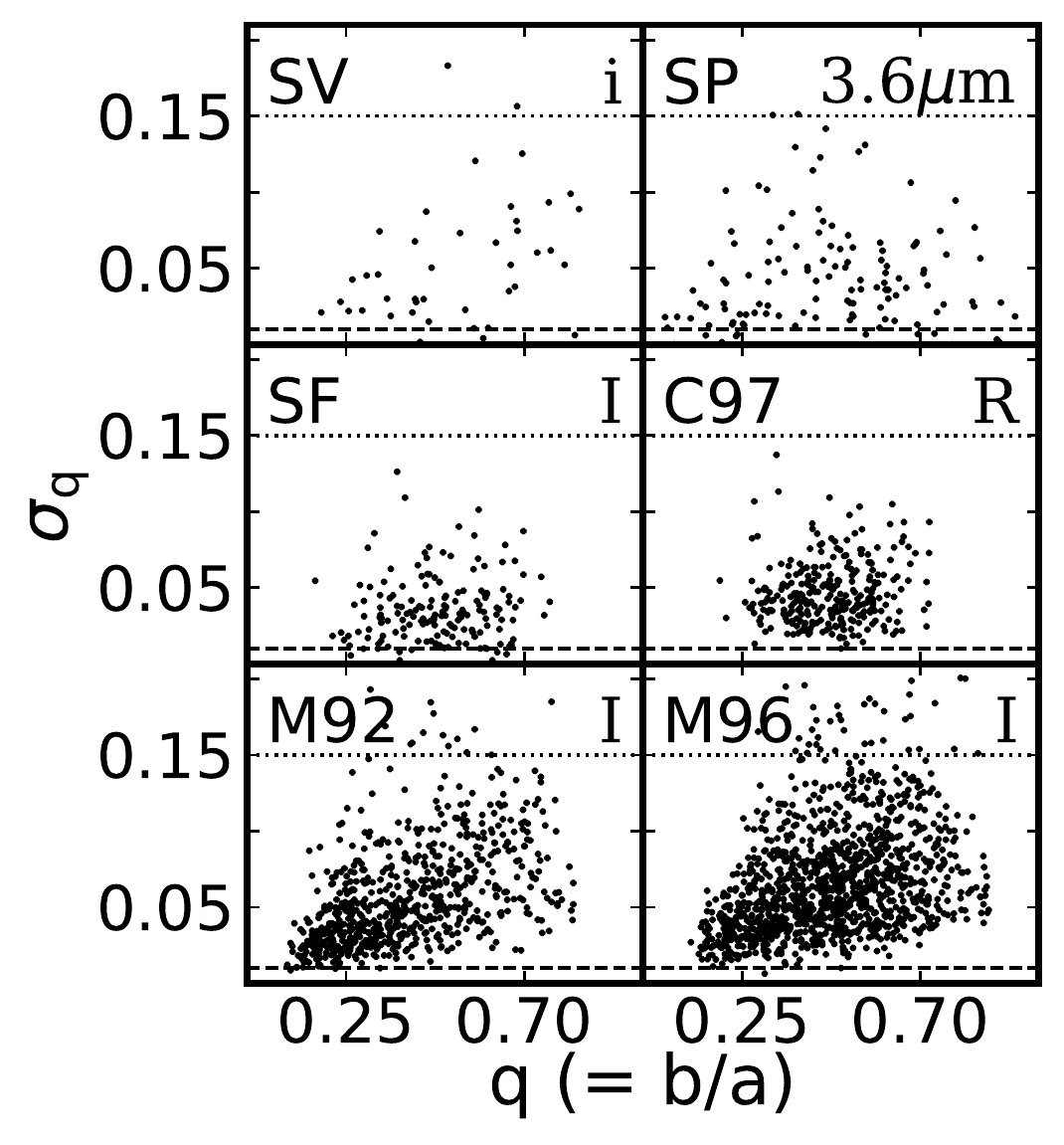}.
Second, the intrinsic flattening parameter $q_0$ is selected with a Gaussian uncertainty, truncated to the range $[0.05-0.25]$, representing typical acceptable values for the intrinsic flattening of a disk~\citep{Kregel2002}.

\begin{figure}
  \centering
  \includegraphics[width=\columnwidth]{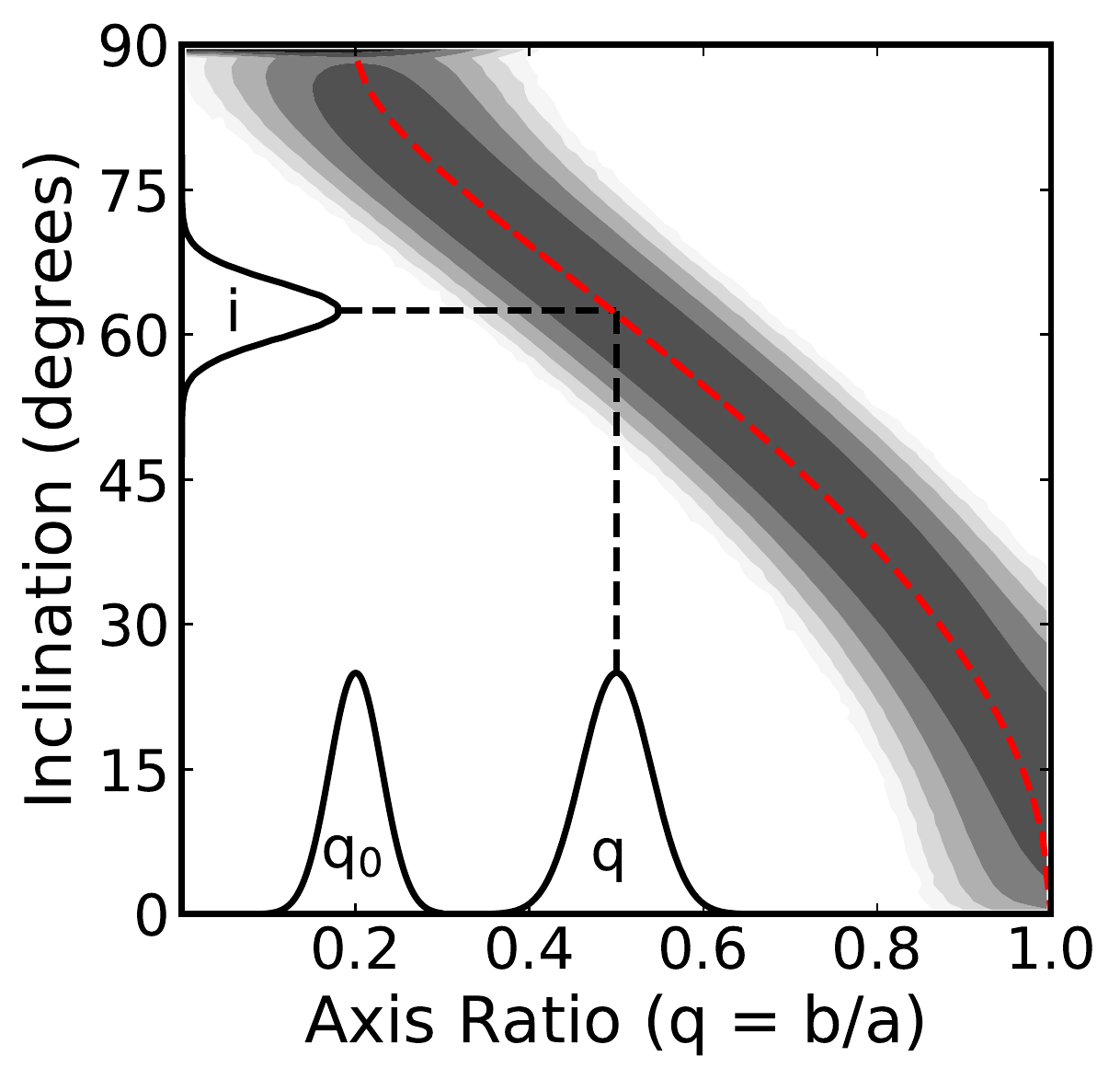}
  \caption{\footnotesize Inclination uncertainty model. The $q$ and $q_0$ distributions shown on the x-axis are folded through \Equ{calcinclination} (red dashed line) in a Monte Carlo simulation to determine the resulting inclination and its uncertainty distribution. The $i$ distribution is taken from the $q$ distribution shown on the x-axis. All $i$ uncertainty distributions can be determined with vertical slices in the contour plot, which has levels chosen evenly in log probability.} 
  \label{fig:Uncertainty_Inclination.pdf}
\end{figure}

These are combined through \Equ{calcinclination} to determine the adopted inclinations.
\Fig{Uncertainty_Inclination.pdf} demonstrates the results of this model.
For a given axis ratio $q$, a vertical slice in \Fig{Uncertainty_Inclination.pdf} gives the corresponding uncertainty distribution in inclination $i$.

\begin{figure}[ht]
  \centering
  \includegraphics[width=\columnwidth]{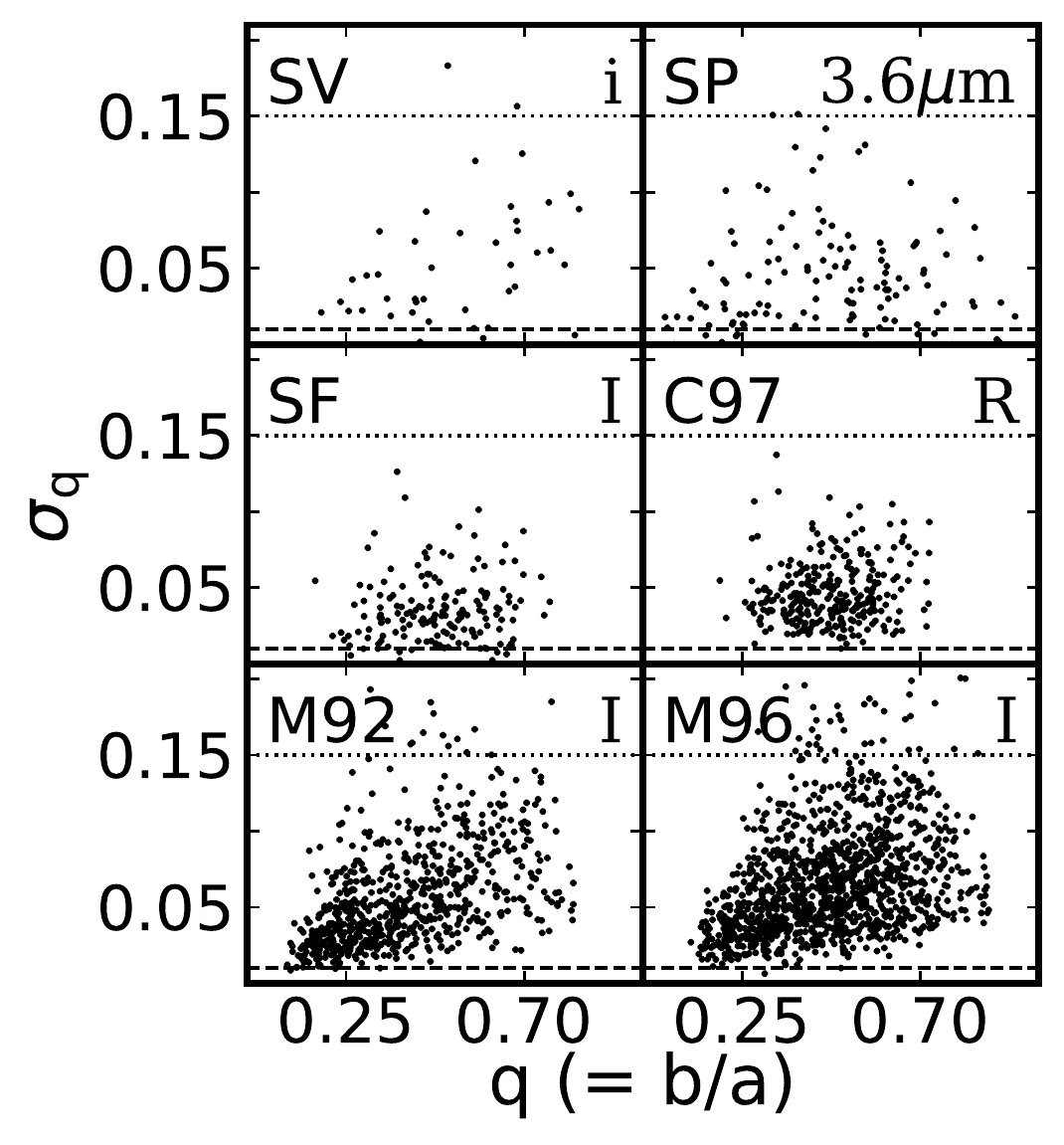}
  \caption{\footnotesize Axis ratio uncertainties for all surveys (as in \Fig{Figures/rar_plot_obs.pdf}) computed from the variation in the isophotal fitting results. The uncertainties fill a similar parameter space for all surveys, with slight differences likely due to instrumentation, wavelength band (shown in the upper right corner), and isophotal fitting technique used. The dashed line represents the minimum allowed uncertainty; any points below that value are truncated at the dashed line. The dotted line represents the quality cutoff threshold for the axis ratio uncertainty.}
  \label{fig:Figures/q_uncertainty.pdf}
\end{figure}

\subsection{Rotational Velocities}
\label{sec:uncertaintyvelocities}

\begin{figure}[ht]
  \centering
  \includegraphics[width=\columnwidth]{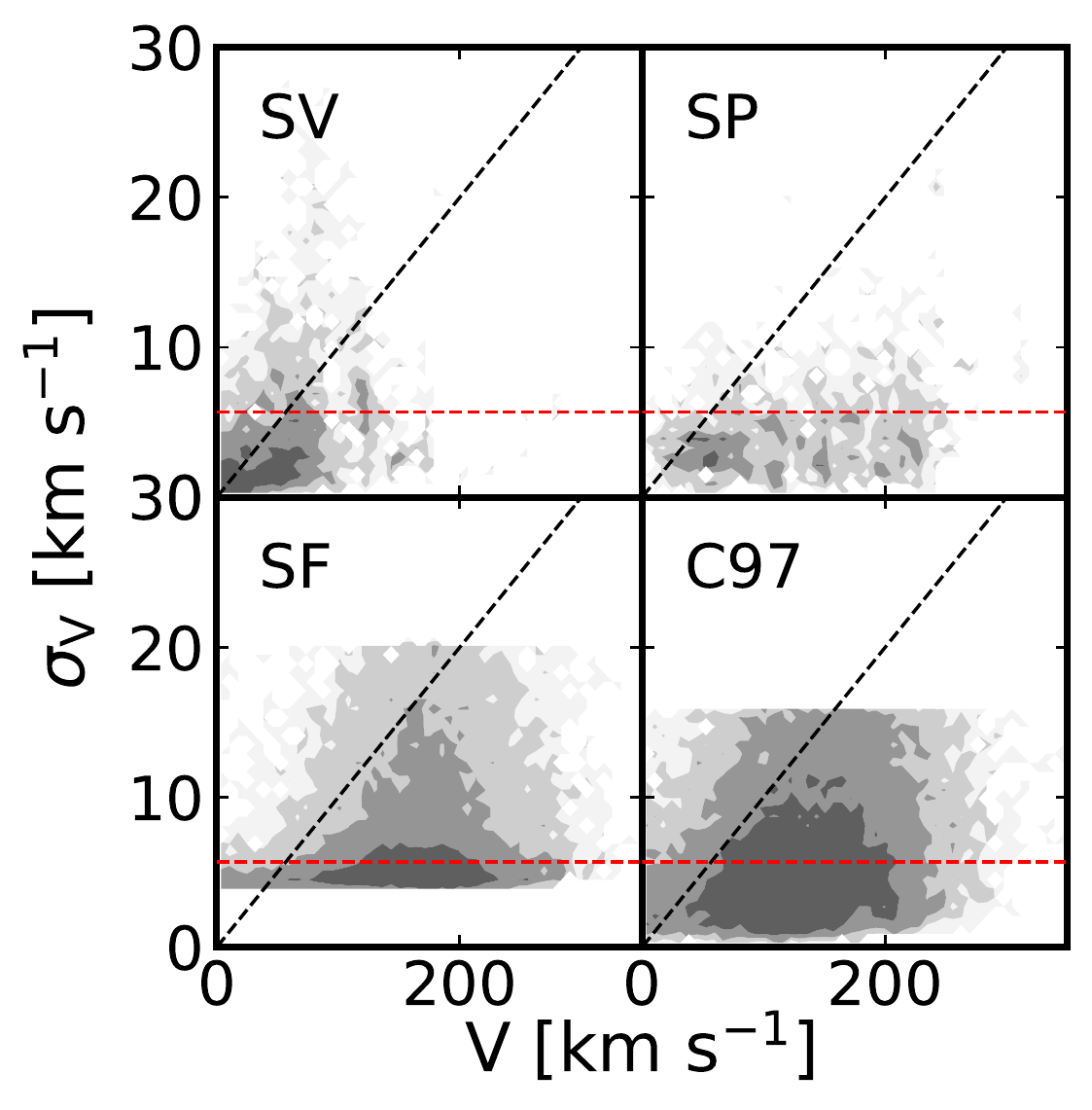}
  \caption{\footnotesize Rotational velocity uncertainties for the 4 surveys (as in \Fig{Figures/rar_plot_obs.pdf}) which report per-point errors. Shown are contour plots with levels evenly spaced in log density. The dashed black line represents the cutoff uncertainty of $\sigma_V/V = 0.1$~(\Lseventeen) above which data points are rejected. SPARC velocities are a mix of \HI and \Ha measurements, while all other surveys use \Ha.} 
  \label{fig:Figures/Uncertainty_Velocity.pdf}
\end{figure}

Rotational velocity uncertainties are reported for each survey except \MATtwo and \MATsix.
For these, a uniform velocity uncertainty of $\sigma_V = \wunits{6}{km\,s$^{-1}$}$ is used, representing the median uncertainty for the other surveys. 
\Fig{Figures/Uncertainty_Velocity.pdf} demonstrates the distribution of relative rotational velocity uncertainties for each survey.
The black dashed line indicates the adopted cutoff threshold of \wunits{10}{\%}~(\Lseventeen), where velocity measurements are not considered in the analysis.
Velocity uncertainties account for measurement errors but ignore noncircular motions (see \Sec{montecarloscatter}).

\subsection{Curves of Growth}
\label{sec:uncertaintymagvalues}

As with rotational velocities, magnitude uncertainties are reported for each survey except \MATtwo and \MATsix.
The reported uncertainties are depicted in \Fig{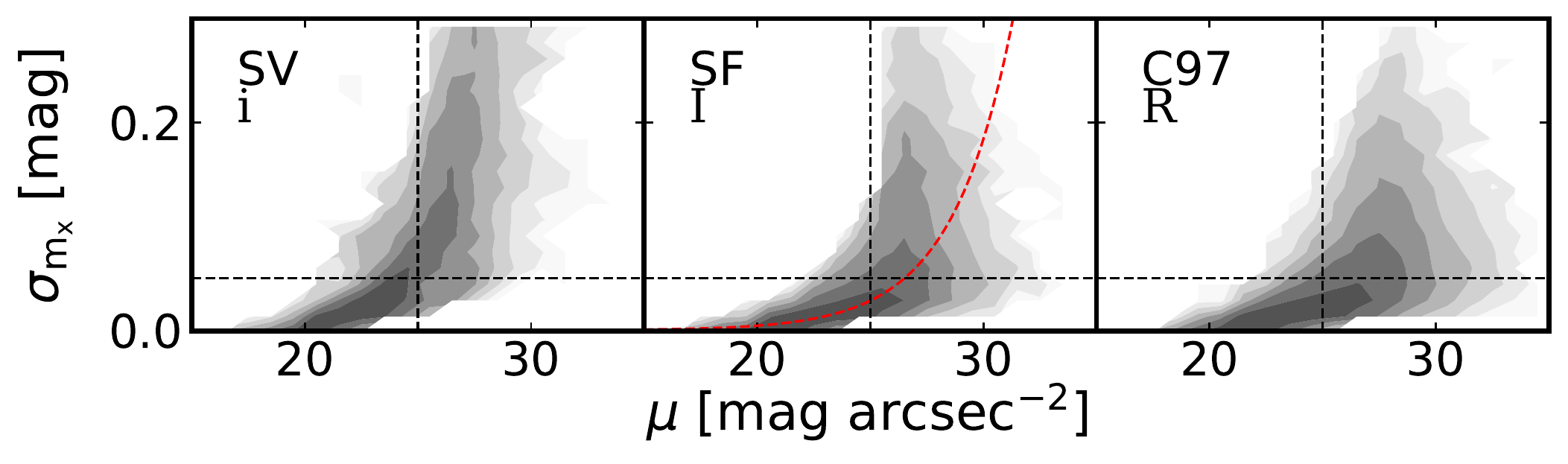} as a function of the SB values.
We find that magnitude uncertainties correlate more tightly with SB than magnitude measurements, hence this choice of variable. 
The dashed red line in the Shellflow panel represents the SB uncertainties used for \MATtwo and \MATsix.
This is done by fitting $\sigma_{\rm SB} = a + e^{b(\mu - c)}$ to the values that meet the data quality threshold (black dashed lines).
The Shellflow distribution is adopted, as it relies on the same photometric band as \MATtwo and \MATsix; 
the resulting parameterization for $(a,b,c)$ is (0.00075, 0.52, 31.97).
Magnitude uncertainties are small compared to other sources of uncertainty; however, they are retained for consistency.

\begin{figure}[ht]
  \centering
  \includegraphics[width=\columnwidth]{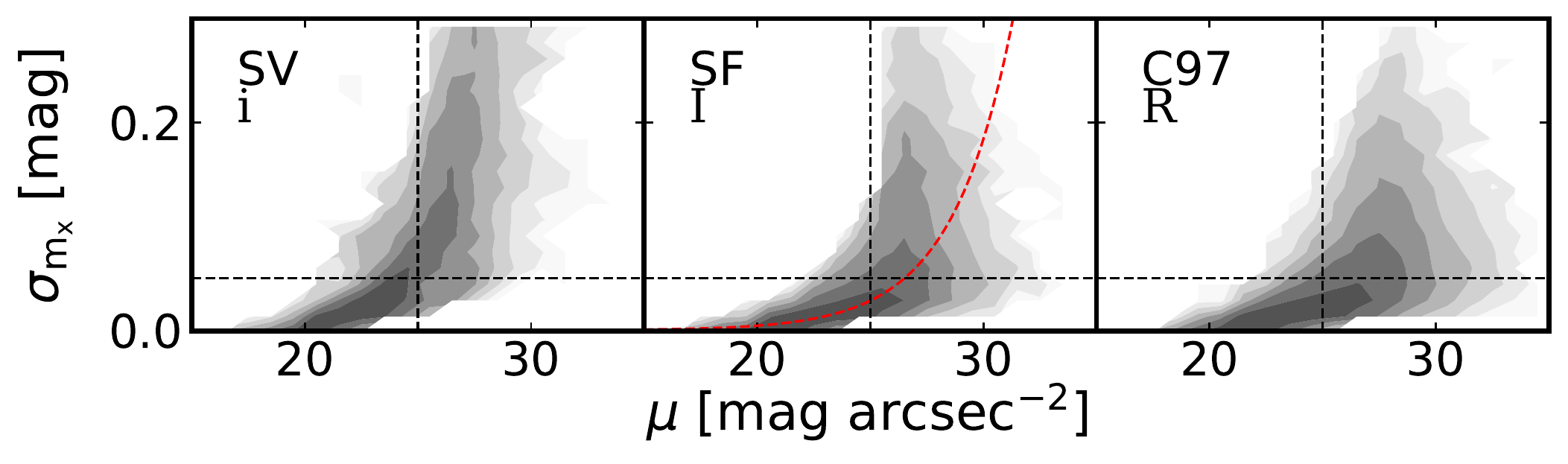}
  \caption{\footnotesize Magnitude uncertainties for the 3 surveys (as in \Fig{Figures/rar_plot_obs.pdf}) reporting per-point errors. The contour plots have levels spaced evenly in log density. The dashed black lines represents the cutoff thresholds for the associated values. Each survey is taken in a different photometric band. The red dashed line in the Shellflow panel represents the adopted magnitude uncertainties in \MATtwo and \MATsix.} 
  \label{fig:Figures/magnitude_uncertainty.pdf}
\end{figure}

\subsection{Stellar Mass-to-light Ratios}
\label{sec:uncertaintymasstolight}

Stellar mass-to-light ratios also carry a significant model uncertainty.
The exact normalization of the mass-to-light ratio is inconsequential to our analysis; only the intrinsic variability about the relation matters.
\citet{Roediger2015} found the random uncertainty of the stellar mass-to-light ratio to be of order \wunits{0.09-0.13}{dex} for optical mass-to-light ratios.
A representative conservative uncertainty of \wunits{0.13}{dex} is therefore used. 
Systematic uncertainties (e.g. model-to-model differences) for the stellar mass-to-light ratio of the order of \wunits{0.3}{dex} \citep{Conroy2013,Courteau2014,Roediger2015} are of course larger; however, this uncertainty is not used for our analysis, as a systematic shift (constant factor in log space) does not impact scatter measurements.
For the uncertainty in the \wunits{3.6}{$\mu$m} stellar mass-to-light ratio, an uncertainty of \wunits{0.11}{dex} is used, as suggested by \citet{Meidt2014}.
The effect of choosing even larger stellar mass-to-light ratio uncertainties is partially explored in \Sec{montecarloscatter}.

\section{Monte Carlo Scatter Model}
\label{sec:montecarloscattermodel}

\subsection{Projecting onto the Stellar RAR}
\label{sec:projectingontotherar}

To assess the scatter induced by observational uncertainties, one must eliminate the effects of intrinsic scatter (if any) in the RAR.
As the observational and intrinsic scatter are mixed, let us first eliminate all scatter.
To this end, all data points in the \gs \gob space are projected onto the fitted RAR to impose a zero-scatter assumption.
To project a data point onto the RAR, its final scatter-free location must first be identified. 
This is equivalent to asking what new values of ${\rm g}_{\rm *,zs}$ and ${\rm g}_{\rm obs,zs}$, where ``zs'' means zero scatter (constrained to the RAR), should be chosen.
One option is to project along the \gs or \gob axis; however, this means artificially selecting an axis.
A second option is to adopt an orthogonal projection, where each ${\rm g}_{\rm *,zs}$ and ${\rm g}_{\rm obs,zs}$ is the minimum distance (in \gs, \gob space) from the measured \gs, \gob.
However, this assumes that the uncertainties in each axis are identical, which may not be true either.
Instead, a third and favored option is to consider the full uncertainty distributions from \Sec{uncertaintymodel} for each parameter, construct a likelihood function by multiplying the probabilities, and choose the location that maximizes the likelihood whilst imposing a scatter-free RAR (see \Equ{rar}).

As every parameter is given a probability distribution in \Sec{uncertaintymodel}, it is possible to construct a likelihood that is a function of each variable but constrained to lay on the RAR.
This removes one degree of freedom for each point on the RAR.
The equations from \Secs{gobs}{gbar} can be used to connect \gob and \gs to their respective measurements,

\begin{eqnarray}\label{equ:constraintsV}
  V = \sin(i)\sqrt{\gob\theta D} \\ \label{equ:constraintsL}
  L_x = \frac{\gs (\theta D)^2}{G\Upsilon_x},
\end{eqnarray}

\noindent where \gob is determined using \Equ{rar}, thus constraining it to lay on the RAR.
Some of the variables are shared between multiple observations in the same galaxy (distance, axis ratio, intrinsic thickness, and mass-to-light ratio), and so the optimization must be performed for all observations simultaneously.
For a galaxy with $N$ observations on the RAR, the likelihood function can be represented as

\begin{eqnarray}\label{equ:zslikelihood}
  &\mathcal{L}({\rm g}_{\rm *,1}', {\rm g}_{\rm *,2}',...,{\rm g}_{\rm *,N}',q',q_0',\Upsilon_x',D') = \\
  &P(q'|q,\sigma q)\cdot P(q_0'|q_0,\sigma q_0)\cdot P(\Upsilon_x'|\Upsilon_x,\sigma\Upsilon_x)\cdot \nonumber\\
  & P(D'|D,\sigma D)\cdot\Pi_{i=1}^N P(V'_i|V,\sigma V)\cdot P(L'_i|L_i,\sigma L_i),\nonumber
\end{eqnarray}

where the primed quantities are the new zero-scatter values and the unprimed quantities are the measurements and their uncertainties.
Here $V_i'$ and $L_i'$ are computed from \Equs{constraintsV}{constraintsL} respectively, using the primed quantities; \gob is computed from \Equ{rar}.
Ultimately, this process finds the values for all quantities that are most consistent with the original measurements while remaining constrained on the RAR.
The values for each variable are now taken as the ``true'' values, upon which the observational uncertainties can scatter measurements off the RAR.
This process is described in the next section.

\subsection{Reintroducing Observational Uncertainties}
\label{sec:reintroducingobservationaluncertainties}

\begin{figure}[ht]
  \centering
  \includegraphics[width=\columnwidth]{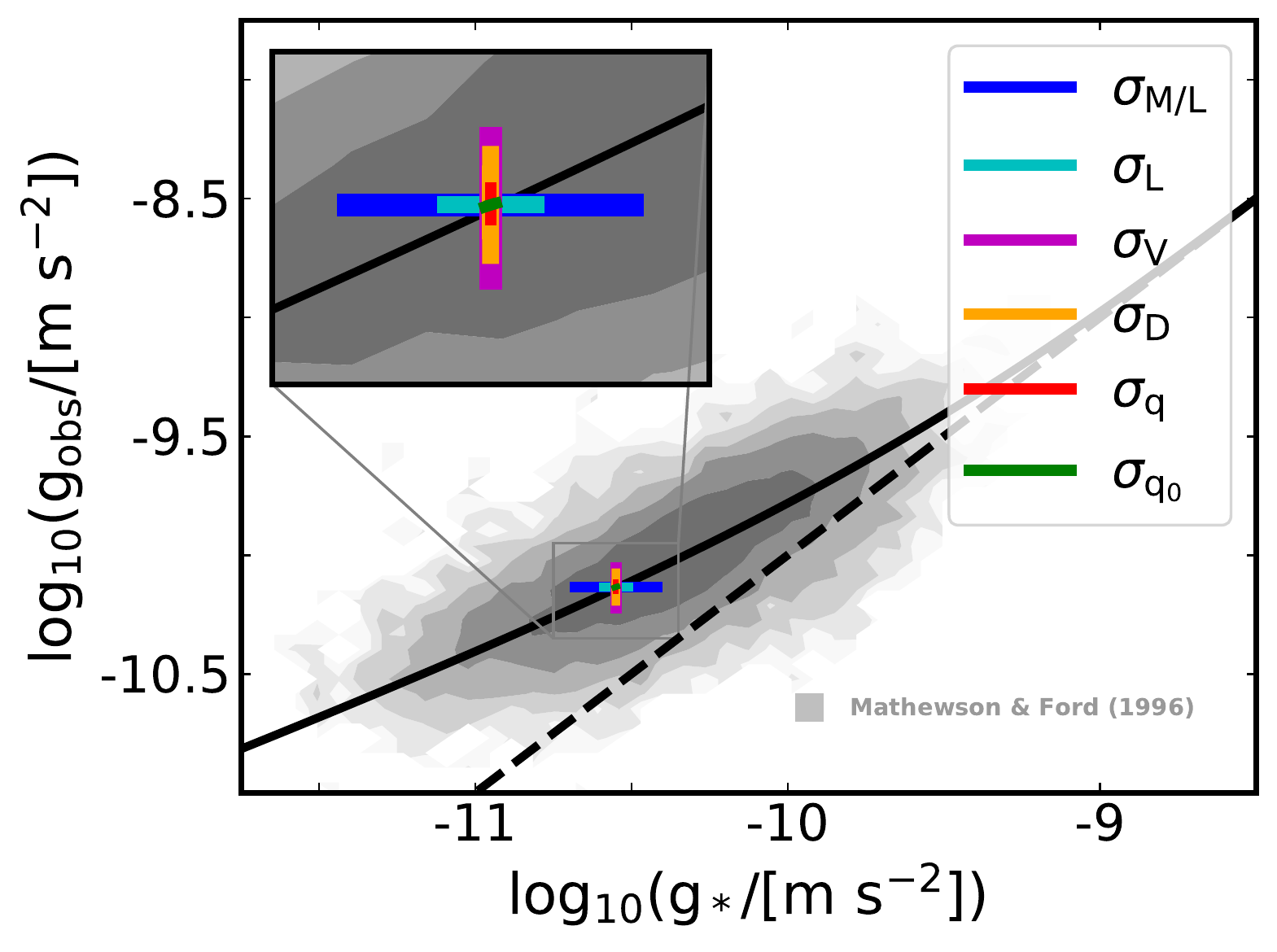}
  \caption{\footnotesize Representation of the variations introduced by each observed variable in the RAR calculation. For each parameter, $\pm 1\sigma$ variations are shown as colored bars radiating from a central point laying exactly on the RAR. For reference, the grey distribution in the background is from the \MATsix survey.}
  \label{fig:Figures/Variations.pdf}
\end{figure}

Using the uncertainty models from \Sec{uncertaintymodel}, new ``observations'' can be sampled about the zero-scatter values as described in \Sec{projectingontotherar}.
To visualize the effect of this Monte Carlo model, \Fig{Figures/Variations.pdf} shows how the different uncertainties affect the final location of a single point off of the RAR.
New data points are sampled for each galaxy and compiled into the same data structure as the original measurements.
As the RAR is a local scaling relation, each galaxy places multiple data points on the relation; however, some variables are global for the whole galaxy.
For example, a single distance value is resampled for each galaxy.
Thus, it is clear that the data points in the RAR are not independent, as multiple measured values with large uncertainties are shared.

To match the statistics of the original observations, each point is resampled only once.
Thus, the simulated data have exactly the same number of initial entries as the measured data.
To include the effects of data quality cuts, all measured values are resampled before the data quality cuts are applied.
This means that some points in the RAR that were initially cut may be newly introduced, for example, if a galaxy had an inclination below $30^{\circ}$ but was resampled above that value.

The mock data are then processed with the same code as the original data, thus duplicating any data quality cuts, interpolations, and conditional statements.
This critical element is missing from a first-order analysis (e.g. \Lseventeen) and can potentially impact the final result, as will be seen in \Sec{montecarloscatter}.
An element of \Lseventeen's analysis that is not reproduced here is the flat-disk model for the baryonic acceleration.
Inverting Poisson's equation adds unnecessary complexity without impacting the scatter measurements that are the primary focus of this paper.
In \Sec{projectingontotherar}, the projection onto the RAR is made under the assumption that the galaxies are spherical; thus, the calculations remain internally consistent.
In both the flat-disk analysis and spherical assumption, the baryonic acceleration depends linearly on luminosity and mass-to-light ratio, which are the two dominant sources of uncertainty in \gb.
The uncertainty model for these quantities should behave identically for the two baryon distributions.

\section{Results}
\label{sec:results}

The PROBES catalog provides a consistent set of data quality cuts and a detailed uncertainty model, enabling a robust characterization of the intrinsic scatter in the RAR.
In order to do so, we perform a first-order analysis of all uncertainties using a simplified error model as in \Lseventeen.
This model incorporates the same uncertainties as our Monte Carlo model (\Sec{montecarloscattermodel}), but only treated to first order.
Next, we consider the full uncertainty model with Monte Carlo sampling to get an accurate measure of the observational uncertainties.
The first-order analysis and the full uncertainty model can then be used to extract the intrinsic scatter from the data; the results are finally compared for each method.

\subsection{First-Order Scatter Prediction}
\label{sec:firstorderscatter}

A first-order calculation of the average observational uncertainty for the SPARC data set was presented by \Lseventeen.
Here we reproduce this calculation for the full PROBES sample. 
\Equs{uncertaintygobs1storder}{uncertaintygbar1storder} present the propagation of uncertainties through the calculation of \gs and \gob.
In \Lseventeen the assumed luminosity uncertainty is $\sigma_{L} = \wunits{0.04}{dex}$, and the assumed mass-to-light ratio uncertainty is $\sigma_{\Upsilon} = \wunits{0.1}{dex}$, meaning that the baryonic acceleration uncertainty is always $\sigma_{\gb}/\gb = \wunits{0.11}{dex}$.
In this paper, we consider the same luminosity uncertainty and more conservative values for the mass-to-light ratio uncertainty of $\sigma_{\Upsilon}^{[3.6\mu m]} = \wunits{0.11}{dex}$ \citep{Meidt2014}.
For optical wavelengths, we use $\sigma_{\Upsilon}^{\rm optical} = \wunits{0.13}{dex}$~\citep{Roediger2015, Zhang2017}.
While both \gs and \gob contribute to the scatter in the forward direction, the stellar acceleration uncertainty is modified by the RAR as the slope changes across the relation.
The average scatter in the forward residuals takes the form

\begin{equation}\label{equ:firstorderscattercalc}
  \sigma_{\rm 1^{st} Order} = \frac{1}{N}\sum^N \frac{1}{\gob}\sqrt{\sigma_{\gob}^2 + \left(\frac{\partial\mathcal{F}}{\partial\gs}\sigma_{\gs}\right)^2}~,
\end{equation}

\noindent where $\sigma_{\rm 1^{st} Order} \approx \sigma_{\rm obs}$ is an estimate of the scatter in the residuals due to observational uncertainties, $N$ is the number of observations, and $\mathcal{F}$ is the functional form of the RAR (\Equ{rar}).
\Tab{firstorderscatters} presents the results of this calculation for each survey, where the first column represents the survey, the second column reports the observed scatter of the RAR, the third column gives the first-order scatter result from \Equ{firstorderscattercalc}, and the fourth column reports the intrinsic scatter resulting from the difference in quadrature ($\sigma_{\rm int}^2 =  \sigma_{\rm RAR}^2 - \sigma_{\rm 1^{st} Order}^2$).
Each survey does indeed have different observational uncertainties, as expected for surveys with such a range of properties.
The median value for the intrinsic scatter is then \wunits{0.13}{dex}; this can be compared to the results found in $\Lambda$CDM simulations, which range from 0.06 to \wunits{0.08}{dex}~\citep{Keller2017,Ludlow2017}.
The value determined from first-order calculations is substantially larger than expected from $\Lambda$CDM simulations and certainly not consistent with zero.

\begin{table}
  \footnotesize
  \begin{center}
    \caption{\\ First-order Scatter Predictions for all PROBES Surveys, with Corresponding Intrinsic Scatters}
    \begin{tabular}{c c c c }
      \tableline\tableline
      1 & 2 & 3 & 4 \\
      Survey & $\sigma_{\rm RAR}$ & $\sigma_{\rm 1^{st} Order}$ & $\sigma_{\rm int}$ \\
      & dex & dex & dex \\ \tableline
      SV & 0.166 & 0.114 & 0.121 \\
      SP & 0.130 & 0.097 & 0.087 \\
      SF & 0.139 & 0.089 & 0.107 \\
      C97 & 0.182 & 0.087 & 0.160 \\
      M92 & 0.173 & 0.092 & 0.146 \\
      M96 & 0.170 & 0.090 & 0.145 \\
      \tableline
    \end{tabular}\label{tab:firstorderscatters}
  \end{center}
  \par {Note. Column (1) indicates the survey (as in \Fig{Figures/rar_plot_obs.pdf}). Column (2) shows the scatter measured from the observed data. Column (3) shows the scatter predictions from a first order analysis. Column (4) shows the intrinsic scatter from the difference of Column (2) and (3) in quadrature.}
\end{table}

The values in \Tab{firstorderscatters} are close to those from \Lseventeen.
Our respective values of $\sigma_{\rm RAR}$ are essentially identical. 
For $\sigma_{\rm 1^{st} Order}$, we find \wunits{0.1}{dex}, while \Lseventeen quoted \wunits{0.12}{dex}.  
This discrepancy originates in slightly different choices in uncertainty values for some parameters.  
We used larger uncertainty values than \Lseventeen for the stellar mass-to-light ratios; for the inclination, we typically found smaller uncertainties with our model from \Sec{uncertaintyinclinations}.
These differences account for most of the \wunits{0.02}{dex} discrepancy.

\subsection{Monte Carlo Simulation Results}
\label{sec:montecarloscatter}

\begin{figure}[ht]
  \centering
  \includegraphics[width=\columnwidth]{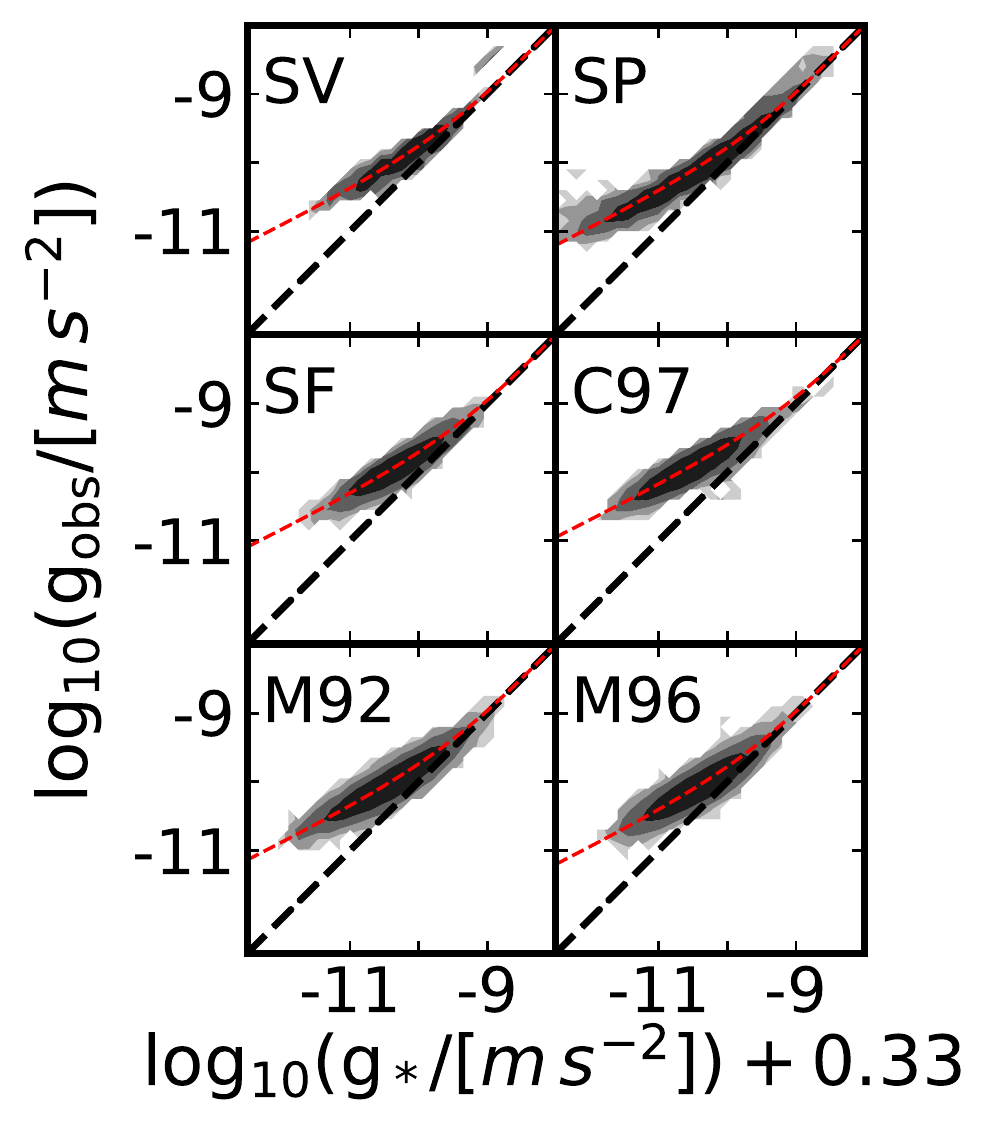}
  \caption{\footnotesize RAR for the mock data (i.e. following the procedure defined in \Sec{montecarloscattermodel}), presented as 2D contours  evenly spaced in log density. The dashed lines represent the one-to-one line and \Equ{rar} fit to the data. Each panel shows a different survey as in \Fig{Figures/rar_plot_obs.pdf}.}
  \label{fig:Figures/rar_plot_sim.pdf}
\end{figure}

Here we present the Monte Carlo simulation of each survey with only observational uncertainties about the RAR.
With the full dataset resampled, any scatter metric (such as the orthogonal scatter) could be chosen to compare with the original data.
For consistency with other studies and with the first-order analysis in \Sec{firstorderscatter}, the forward residuals will be used as the scatter metric.
\Fig{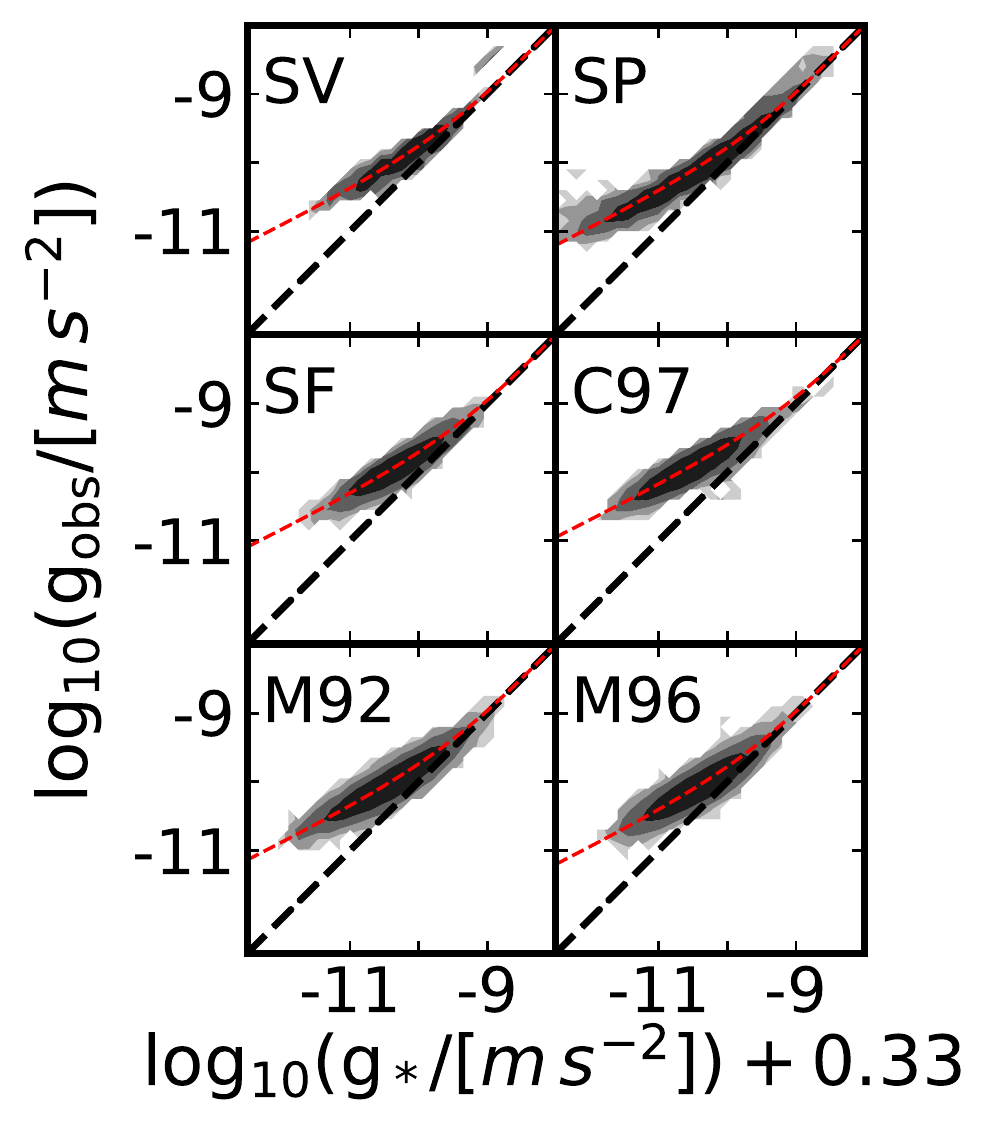} presents the simulated data with the same analysis as in \Sec{fittingoobservedrar}.
Many qualitative elements from \Fig{Figures/rar_plot_obs.pdf} are retained; however, the distributions in \Fig{Figures/rar_plot_sim.pdf} are clearly smoother and tighter.
This visual impression is confirmed in \Tab{montecarlofits} which lists the same parameters as \Tab{observedfits}, where the scatter measured for each simulated survey is below that of the observed one.
\Tab{scatterresults} presents the scatter measurements for the mock data in the same format as \Tab{firstorderscatters}.

\begin{table}
  \footnotesize
  \begin{center}
    \caption{\\ Parameterization for the RAR Fit to Each Mock Data Set.}
    \begin{tabular}{ c  c  c  c }
      \tableline\tableline
      1 & 2 & 3 & 4 \\
      Survey & $\gd$ & $\sigma_{\rm sim}$ & $N$ \\
      & $10^{-10}~m\,s^{-1}$ & dex & \# \\ \tableline
      SV   & $0.768 \pm 0.020$ & $0.134 \pm 0.004$ & 1685 \\
      SP   & $1.051 \pm 0.021$ & $0.125 \pm 0.003$ & 2416 \\
      SF   & $0.611 \pm 0.004$ & $0.117 \pm 0.001$ & 17,173 \\
      C97  & $5.626 \pm 0.024$ & $0.100 \pm 0.001$ & 18,065 \\
      M92  & $1.810 \pm 0.009$ & $0.130 \pm 0.001$ & 18,401 \\
      M96  & $1.340 \pm 0.007$ & $0.121 \pm 0.001$ & 21,137 \\
      \tableline
    \end{tabular} \label{tab:montecarlofits}
  \end{center}
  \par {Note. Column (1) indicates the survey (as in \Fig{Figures/rar_plot_obs.pdf}). Column (2) is the parameterization for \Equ{rar} and its bootstrap uncertainty. Column (3) is the \wunits{16-84}{\%} interval scatter reported with its bootstrap uncertainty. Column (4) gives the number of data points in the RAR.}
\end{table}

As with our first-order scatter calculations in \Sec{firstorderscatter}, one may compute the intrinsic scatter by taking the difference in quadrature with the observed scatter.
Using the mock data (\Tab{scatterresults}), the median intrinsic scatter is $\sigmaint \pm 0.02$, where the uncertainty is the weighted (by points in the RAR) standard deviation of the intrinsic scatter values.
These results are in closer agreement with $\Lambda$CDM expectations \citep{Ludlow2017,Keller2017,Dutton2019} than the first-order analysis.
\citet{Dutton2019} noted that the baryonic RAR scatter depends on galaxy mass.
We observe the same effect with PROBES galaxies but not our mock data.
\App{massbins} explores the RAR as a function of mass bins.
This sheds light on the range of scatter measurements amongst surveys that sample different mass regimes.

\begin{table}
  \footnotesize
  \begin{center}
    \caption{\\ Scatter measurements for the RAR.}
    \begin{tabular}{ c  c  c  c }
      \tableline\tableline
      1 & 2 & 3 & 4 \\
      Survey & $\sigma_{\rm RAR}$ & $\sigma_{\rm sim}$ & $\sigma_{\rm int}$ \\ 
      & (dex) & (dex) & (dex) \\ \tableline
      SV   & $0.166 \pm 0.007$ & $0.134 \pm 0.004$ & $0.098 \pm 0.013$ \\
      SP   & $0.131 \pm 0.003$ & $0.125 \pm 0.003$ & $0.040 \pm 0.013$ \\
      SF   & $0.139 \pm 0.001$ & $0.117 \pm 0.001$ & $0.076 \pm 0.002$ \\
      C97  & $0.182 \pm 0.001$ & $0.100 \pm 0.001$ & $0.153 \pm 0.002$ \\
      M92  & $0.173 \pm 0.001$ & $0.130 \pm 0.001$ & $0.114 \pm 0.002$ \\
      M96  & $0.170 \pm 0.001$ & $0.121 \pm 0.001$ & $0.120 \pm 0.002$ \\
      \tableline
    \end{tabular} \label{tab:scatterresults}
  \end{center}
  \par {Note. Column (1) indicates the survey (as in \Fig{Figures/rar_plot_obs.pdf}). Column (2) reports the scatter in the observed data about the fits presented in \Tab{observedfits}, Column (3) reports the scatter in the Monte Carlo data about the fits in \Tab{montecarlofits}. Column (4) shows the intrinsic scatter calculated by subtracting $\sigma_{\rm RAR}$ from $\sigma_{\rm sim}$ in quadrature.}
\end{table}

Several factors may potentially cause the Monte Carlo simulated scatter to underrepresent the full observational uncertainties.
For instance, noncircular motions, which might increase the observed scatter, especially in the inner disk, are not modeled here.
We find that the median intrinsic scatter for RAR points beyond one $R_{e}$, i.e. for regions relatively free of noncircular motions, is \wunits{0.10\pm0.01}{dex}, still in excellent agreement with $\Lambda$CDM simulations~\citep{Ludlow2017,Keller2017,Dutton2019}.

It is also possible that some of the observational uncertainties have been underestimated.
To assess that possibility, we consider a scaling factor that is applied to all uncertainty values for a given parameter.
We consider one parameter $x$ at a time, with its uncertainty being scaled as $\sigma_x' = \epsilon \sigma_x$ and $\epsilon$ as the scale factor, whilst all the other uncertainties are kept at their fiducial value.
The scaling factor is tuned until the Monte Carlo simulated scatter is in agreement with the observed scatter (indicating zero intrinsic scatter).
The final scaling factor value is reported in \Tab{uncertaintyscalefactors}.
 
This exercise suggests that, in order to explain away all intrinsic scatter in the RAR, the parameter errors would need to be inflated by factors of $2$ (stellar mass-to-light ratio) to $10$ (total luminosity and disk flattening ratio).  
Considering the already conservative values adopted for the uncertainty in each variable, it seems unlikely that the scaled uncertainties represent a reasonable path to explain the nonzero scatter.
We reiterate that the nonzero scatter is most likely a true feature of the data, in agreement with $\Lambda$CDM simulations. 

\begin{table}
  \footnotesize
  \begin{center}
    \caption{\\ Scaling factors for the uncertainty in each observed parameter.}
    \begin{tabular}{ c  c  c  c  c  c  c  c }
      \tableline\tableline
      Parameter & $V$ & $\Upsilon$ & $L$ & $q$ & $q_0$ & $D$ & $m_0$ \\ \tableline
      $\epsilon$ & 3.0 & 1.7 & $>10$ & 7 & $>10$ & 2.1 & 5 \\
      \tableline
    \end{tabular}\label{tab:uncertaintyscalefactors}
  \end{center}
  \par {Note. When scaled by one of these factors, the uncertainties reported here would be large enough to produce a median intrinsic scatter equal to zero. The parameters (L to R) are rotational velocity, stellar mass-to-light ratio, total luminosity, disk axial ratio, intrinsic disk flattening, distance, and magnitude zero-point.}
\end{table}

\section{Conclusions}
\label{sec:conclusions}

We have presented a method for determining the intrinsic scatter about any scaling relation using the stellar RAR as a test case.
The database used in this paper represent the largest collection of extended spatially resolved RCs and SB profiles to date.
With over 2500 galaxies sampled over a representative range of galaxy properties, the detailed study of numerous scaling relations such as the stellar RAR becomes possible.

As demonstrated in \Msixteen, the baryonic RAR presents an interesting local scaling relation with the capacity to test alternative dark matter models.
We have examined the intrinsic scatter in the stellar RAR with great care in order to accurately model observational uncertainties.
All six PROBES galaxy surveys, each with different selection criteria, wavelength coverage, and instrumental setups, were analyzed simultaneously to minimize systematic biases.
Our Monte Carlo modeling of the observational uncertainties enables a simultaneous treatment of all nonlinear effects.
Ultimately, we find that the intrinsic scatter in the stellar RAR is of order \wunits{\sigmaint \pm 0.02}{dex}, which broadly agrees with $\Lambda$CDM simulations that predict \wunits{0.06-0.08}{dex}~\citep{Keller2017,Ludlow2017}.
In order to explain away any intrinsic scatter that we ascribe to $\Lambda$CDM effects, the observational uncertainties would have to be considerably inflated from their already conservative estimates (\Tab{uncertaintyscalefactors}). 
Ultimately, our conservative assessment of errors in the stellar RAR yields full consistency with $\Lambda$CDM expectations.

\acknowledgments

We are grateful to the Natural Sciences and Engineering Research Council of Canada, Ontario Government, and Queen's University for support through various scholarships and grants.
The referee is thanked for a thoughtful and constructive report.
Nikhil Arora and Larry Widrow are also thanked for illuminating discussions.
The software packages {\it galpy} and {\it astroquery} were used extensively in this analysis.  
We also acknowledge the NASA/IPAC Extragalactic Database for the wealth of information that it provides. 

\appendix

\section{Inverse RAR Scatter}\label{app:inversescatter}

We have presented the RAR  with \gs (the stellar mass-dependent quantity) on the x-axis and \gob (the velocity-dependent quantity) on the y-axis.
This choice impacts  scatter measurements, which have so far been measured from forward (vertical) residuals.
Inverse scatter is measured horizontally in the RAR diagram.   
Rather than measuring the scatter of \gs for a given \gob (forward method), we now compute the inverse scatter of \gob for a given \gs (inverse method).
At high acceleration, the slope of \gob versus \gs is roughly 1, meaning the forward and inverse scatters are the same. However, at low acceleration, the slope of \gob versus \gs is $\sim$0.5; thus, the forward scatter is about half that of the inverse relation. 

\Tab{invscatter} shows the inverse scatter calculated using the same techniques as in \Tab{scatterresults}.
In some cases (e.g. Shellflow), the simulated scatter can be slightly larger than the observed scatter; this unphysical result simply results from the statistical nature of our analysis, and the intrinsic scatter is consistent with zero.

\begin{table}[ht]
  \footnotesize
  \begin{center}
    \caption{\\ Inverse RAR scatter measurements.}
    \begin{tabular}{ c  c  c  c  }
      \tableline\tableline
      1 & 2 & 3 & 4 \\
      Survey & $\sigma_{\rm RAR}^i$ & $\sigma_{\rm sim}^i$ & $\sigma_{\rm int}^i$ \\ 
      & (dex) & (dex) & (dex) \\ \tableline
      SV   & $0.276 \pm 0.011$ & $0.188 \pm 0.006$ & $0.202 \pm 0.016$ \\
      SP   & $0.229 \pm 0.005$ & $0.184 \pm 0.004$ & $0.137 \pm 0.010$ \\
      SF   & $0.175 \pm 0.002$ & $0.177 \pm 0.001$ & $\llap{-}0.027 \pm 0.014$ \\
      C97  & $0.251 \pm 0.002$ & $0.176 \pm 0.001$ & $0.180 \pm 0.003$ \\
      M92  & $0.277 \pm 0.002$ & $0.207 \pm 0.002$ & $0.184 \pm 0.004$ \\
      M96  & $0.237 \pm 0.002$ & $0.192 \pm 0.001$ & $0.140 \pm 0.003$ \\
      \tableline
    \end{tabular}\label{tab:invscatter}
  \end{center}
  \par {Note. Column (1) shows the survey acronym (as in \Fig{Figures/rar_plot_obs.pdf}). Column (2) shows the scatter in the observed data about the fits presented in \Tab{observedfits}, while column (3) shows the scatter in the mock data about the fits in \Tab{montecarlofits}. Column (4) gives the intrinsic scatter calculated by taking the difference of $\sigma_{\rm RAR}^i$ and $\sigma_{\rm sim}^i$ in quadrature. The superscript $i$ indicates that these are inverse scatter measurements.}
\end{table}

The median of all of the intrinsic scatter measurements is $\sigma_{\rm int}^i = \wunits{0.16\pm 0.04}{dex}$, where the uncertainty $\pm 0.04$ is the weighted  standard deviation of the $\sigma_{\rm int}^i$ values.
Thus, we find that the intrinsic scatter value is statistically equal to the forward scatter, albeit with a much larger variance indicating that the scatter is poorly constrained.

Overall, we adopt a median observed forward and inverse scatter of $\sigma_{\rm RAR} = \sigmaobs$ and $\sigma_{\rm RAR}^i = \wunits{0.24}{dex}$, respectively. 
The choice of forward or inverse residuals greatly impacts the resulting scatter.
Needless to say, this consideration affects all galaxy scaling relations.

A comparison between RAR and BTFR scatters is warranted once the different units are accounted for.
\citet{Hall2012} examined the BTFR scatter in both axes, finding $\sigma_{VM_{\rm bar}} = \wunits{0.078}{dex}$ and $\sigma_{M_{\rm bar}V} = \wunits{0.274}{dex}$.
The forward and inverse scatters are again quite different, with the forward scatter being numerically smaller than the baryonic RAR scatter from \Lseventeen.
However, in order to properly compare with the RAR, the forward scatter of the BTFR must be multiplied by ~$\sim 1.5$, since on one axis, the BTFR scales with $V$ and the RAR with $V^2$. The scatter of the pseudo-BTFR is thus $0.078*1.5 = \wunits{0.12}{dex}$, in very close agreement with the baryonic RAR scatter found by \Lseventeen (\wunits{0.11}{dex}).

\section{Mass-dependent Stellar RAR Scatter}\label{app:massbins}

This appendix presents a study of the RAR scatter as a function of mass bins.
\citet{Dutton2019} found that the baryonic RAR scatter decreases with increasing mass, finding a scatter of $\sigma_{\rm RAR,\Lambda CDM} = \wunits{0.11}{dex}$ in the $M_{*}$ range [$10^{7},10^{9.3}$] $M_{\odot}$, and $\sigma_{\rm RAR,\Lambda CDM} = \wunits{0.04}{dex}$ in the range [$10^{9.3},10^{11}$] $M_{\odot}$.
A number of factors may explain this trend.
Firstly, as noted in \citet{Dutton2019}, the low- and high-mass galaxies exhibit different RC shapes, and the extra scatter at low mass is likely a reflection of the diversity of dwarf galaxy RCs~\citep{Oman2015,Oman2019}.
Secondly, covariant errors between the RAR axes (such as with $q_0$ seen in \Fig{Figures/Variations.pdf}) can result in data points moving along the relation in the high-acceleration regime; however, in the low-acceleration regime, where the RAR flattens, these errors boost the scatter.
Thirdly, the higher observed scatter at lower masses also reflects the higher relative velocity errors on the corresponding galaxy RCs.  
Most velocity measurements have errors of order \wunits{3-10}{km\,s$^{-1}$} (\Fig{Figures/Uncertainty_Velocity.pdf}).

\begin{table}[ht]
  \footnotesize
  \begin{center}
    \caption{\\ Scatter measurements for the RAR as a function of mass bin, column (1).}
    \begin{tabular}{ c  c  c  c  c  c  c }
      \tableline\tableline
      1 & 2 & 3 & 4 & 5 & 6 & 7 \\
      Mass Bin & Survey & $\sigma_{\rm  RAR}$ & $\sigma_{\rm sim}$ & $\sigma_{\rm int}$ & $N_{\rm obs}$ & $N_{\rm sim}$ \\
      log$_{10}\left(\frac{M_{\rm dyn}}{M_{\odot}}\right)$ &  & dex & dex & dex & \# & \# \\ \tableline
      \multirow{6}{*}{8-10}& SV & 0.178 & 0.093 & 0.152 & 879 & 822\\ 
      & SP & 0.144 & 0.091 & 0.112 & 376 & 402\\ 
      & SF & - & - & - & 0 & 0\\ 
      & C97 & 0.179 & 0.107 & 0.144 & 357 & 594\\ 
      & M92 & 0.160 & 0.153 & 0.047 & 788 & 1056\\ 
      & M96 & 0.210 & 0.124 & 0.170 & 512 & 829\\ \tableline
      \multirow{6}{*}{10-11}& SV & 0.144 & 0.145 & \llap{-}0.018 & 501 & 862\\ 
      & SP & 0.116 & 0.125 & \llap{-}0.047 & 796 & 768\\ 
      & SF & 0.130 & 0.094 & 0.090 & 2222 & 1329\\ 
      & C97 & 0.193 & 0.100 & 0.164 & 9036 & 8281\\ 
      & M92 & 0.173 & 0.130 & 0.115 & 10,963 & 11,708\\ 
      & M96 & 0.172 & 0.125 & 0.119 & 13,170 & 13,449\\ \tableline
      \multirow{6}{*}{11-13}& SV & - & - & - & 0 & 0\\ 
      & SP & 0.134 & 0.142 & \llap{-}0.048 & 1387 & 1246\\ 
      & SF & 0.138 & 0.119 & 0.070 & 14,426 & 15,844\\ 
      & C97 & 0.172 & 0.095 & 0.143 & 8233 & 9190\\ 
      & M92 & 0.161 & 0.124 & 0.102 & 6604 & 5637\\ 
      & M96 & 0.147 & 0.113 & 0.095 & 7778 & 6859\\ \tableline      
    \end{tabular}\label{tab:massdeptscatter}
  \end{center}
  \par {Note. Column (2) shows the PROBES survey (as in \Fig{Figures/rar_plot_obs.pdf}), Column (3) shows the scatter in the observed data about the fits presented in \Tab{observedfits}, and Column (4) shows the scatter in the mock data about the fits in \Tab{montecarlofits}. Column (5) gives the intrinsic scatter calculated by taking the difference in quadrature of columns 3 and 4. Columns (6) and (7) give the number of data points used in the calculation.}
\end{table}

\noindent Since high- and low-mass galaxies have peak velocities around 200 and 50~{km\,s$^{-1}$}, respectively, the relative velocity (and thus acceleration) errors are higher for low-mass galaxies.

We can quantify the mass dependence of the RAR scatter and use the mock data to determine if the dependence is intrinsic.
\Tab{massdeptscatter} addresses this issue by examining three dynamical mass bins, where the total mass is computed within the maximal extent of each RC (typically near $R_{23.5}$).
Some of the measurements produce negative intrinsic scatter values (e.g. for SHIVir and SPARC).  These are cases where our conservatively large uncertainty estimates break down, and the negative intrinsic scatter can be viewed as statistical fluctuations.

A clear trend of decreasing scatter with increasing mass is detected for the observed data; for these, the median values of the RAR scatter for each bin are 0.178, 0.158, and 0.147, respectively.
No clear trend is detected for the simulated data where, the median scatter values per mass bin are 0.107, 0.125, and 0.119, respectively.
Thus, the intrinsic scatter, calculated by subtracting the observed scatter from the simulated scatter in quadrature, shows a distinct dependence on mass with median values of 0.144, 0.103, and 0.095, respectively.
We conclude that there is a clear trend of decreasing intrinsic scatter for the observed RAR as a function of mass, independent of signal-to-noise variations, likely bolstering the case for enhanced RC diversity at low masses.

\bibliography{Articles,arXiv,Books,ReviewArticles}

\end{document}